# From Friedel oscillations and Kondo effect to the pseudogap in cuprates

H. Alloul

Laboratoire de Physique des Solides, UMR CNRS 8502, Université Paris-Sud 11, 91405, Orsay (France)

**Abstract.** One of the great achievements performed by Jacques Friedel in France has been to promote experimental activities on the electronic properties in condensed matter physics, which has led him, among others, to be an active promoter of the up-rise of the "Laboratoire de Physique des Solides" (LPS). I have had the chance to be involved in this process in the early days and to create a Nuclear Magnetic Resonance (NMR) group which remains active nowadays. Among various other scientific activities, I have performed some work on Kondo effect in the 1970's which has allowed me to enlighten the real interest of the NMR probe as a local experimental technique to visualize the perturbation brought by impurities in normal metals. These experiments were of course driven by my strong exposure to J. Friedel's education on charge and spin density oscillations in metals. This work has been also my first personal contact with correlated electron physics, a field in which I have been involved for years particularly since the discovery of high $T_c$ cuprates (HTSC). I recall that the early NMR experiments done in these systems have given evidence for the existence of strong electronic correlations and established the occurrence of the "pseudogap", on which I have initially had some exchanges with J. Friedel. The pseudogap, occurring below a temperature $T^*$, has during the last twenty years raised endless discussions about the incidence of correlations on superconductivity. Many researchers are still advocating today that the pseudogap is due to the existence of preformed pairs above $T_c$. I shall show that its robustness to impurities and disorder that we evidenced from the early days rather suggested that it results from a competing order of the correlated state. This is apparently better accepted nowadays, although various distinct ordered states detected below $T^*$ are not yet understood. The short reviews of the work done on Kondo effect and on the cuprates in Orsay allow me to enlighten the importance of the initial choices done by J. Friedel for the LPS. I also underline the large contrast between the scientific behaviours which prevailed between these two periods of scientific activity. This is not only ascribed to the discovery of the HTSC but also to the changes in publication policies and to the modification of evaluation procedures, all this being driven by the advent of information technologies. I suggest that these changes might have a negative incidence, in my opinion, on the research and education system, at least in our field of science.

## 1 Introduction

Most of the French scientists who have been working from 1950 to these days on condensed matter physics-related problems, either in universities or in industrial research laboratories in France have been influenced by J. Friedel's knowledge and views about a broad range of aspects covering fundamental research, metallurgy, materials, scientific education, etc… So the list of French individuals who could write in a Festschrift in honour of the ninetieth birthday of J. Friedel would by itself nearly fill a book. I have therefore been honoured to be selected by an international panel of physicists to write a paper in this Festschrift upon research fields in which J. Friedel has participated actively, that is, Magnetism and Superconductivity. I take that as recognition of the importance of the work I have done on High Temperature Superconductors during the last twenty years with my latest students and collaborators.

I assume also that, as I have been working for so many years, until today, in the "Laboratoire de Physique des Solides" (LPS) created by J. Friedel, I am considered, together with D. Jérome, as one of the major French experimental contributors in the field who had been constantly in contact with J. Friedel. There is no doubt that J. Friedel has always considered that the creation of LPS has been an important achievement in his career. This could be testified by any researcher who has gone through the LPS from the 1960's to about 1990, as J. Friedel was really active in the lab, attended many seminars, PhD committees and administrative and scientific meetings within the university and in the LPS, that he directed for a while. Not a single visitor or member of the LPS at that time could pretend that he has not been influenced in some way or another by J. Friedel.

This honour to contribute to this Festschrift obviously has a cost and I wondered whether I would be able to produce a significant contribution, especially within the short given delay. An easy way would have been to produce a selected summary of one of my recent research activities, either on HTSC, fullerites or cobaltate compounds. But, as J. Friedel had not been involved in those, this would have hardly allowed me to illustrate his impact and different aspects of his personality. So I decided to privilege older work done at the time he was very active at the LPS.

I never co-authored any paper with J. Friedel, but that is by far not a measure of his scientific impact as, contrary to many other scientists, he never abused of forced co-authoring, and I learned from him not do so either. Looking at my publications, I noticed that we often acknowledged discussions with him, which is a good indication that we got exchanges about our experimental results before publication. This has been the case for work I have done on Kondo impurities in noble metals and in the first year of my activity on the cuprates, just after the HTSC up-rise, that is in 1989.

As everybody in the community knows of course about Friedel's oscillations, phase shifts and the relationship between screening and scattering he has promoted in the 1950's [1], it is clear that this was a great part of his teaching and so it appears natural to see members of his laboratory motivated by the Kondo effect. After 1989, J. Friedel did not appear often anymore at the LPS, and I scarcely discussed with him of our HTSC work. But of course the education I got had an incidence on the achievements of my research group.

So I started this essay, focussing on these two topics, which happen furthermore to be relevant for the present Festschrift. Gathering my recollections, I noticed the sharp "sociological" differences which characterized these two scientific periods, so I decided to cover at least partly this aspect together with illustrating the incidence of J. Friedel's actions. I think that, this being personal, the readers will find some unexpected views. Many aspects will be related with scientific education, materials and the relationship between experiment and theory. I therefore found it legitimate to start then this essay with some indications on the events which led me to become J. Friedel's student.

To help readers who are not aware of the education system in France, I briefly recall in §2 my origins, the specificities of the French cursus I followed up to my PhD and I describe briefly the creation of the LPS. Then I discuss in §3 how I have been naturally led to work on spin density oscillations induced by local moment impurities in noble metals and on Kondo effect. I found that recent developments justify reporting here some unpublished results on the matter. I shall then discuss in §4 the HTSC discovery and some results obtained in my group. Most of the scientific debates opened by this discovery certainly underline the richness of the problems raised by electronic correlations in condensed matter physics and have induced a pronounced scientific evolution, if not a revolution. I however consider that, cumulated with the fast outcome of information technologies, this has generated a complete modification of scientific behaviour in our international community, that will be commented in §5. Those aspects are hardly touched by scientists except in private discussions, but have profound impacts on publication policies and scientific careers. This will give me an opportunity to do some final remarks about science education in our field of research, for which my own contributions have been greatly influenced by J. Friedel's views.

## 2  From my discovery of scientific research to the genesis of the LPS

A conjunction of events linked with my origin and youth have unexpectedly driven me to become a physics student. I shall discuss first how that was possible within the education system which applied in the 1960's in France. It had led me one day to meet J. Friedel and to undertake a scientific experimental activity in condensed matter physics.

### 2.1 From elementary to graduate education

Let me recall first that, being born in Marrakech in a Moroccan Jewish family, I never had any contact in my childhood with scientific research, universities or superior education. At the end of high school my math professor suggested that I had the ability to pass the entrance examination to a French "Grande Ecole d'Ingénieurs". I just considered that this was a good idea which could be an opportunity for me to do something practical, work in a factory or construct bridges or anything like that… Then my path was drawn, and I went to the two years preparatory courses for the entrance examinations. The nearest place at that time was in Casablanca, but there my professors told me that going to Paris for the second year would give me a better chance to enter Polytechnique.

I did so successfully. I became a student there and was exposed to a heavy cursus, with no allowed choice, which permitted to open the student's eyes on many subjects in a short-while, quite different from the university system for which you have to be acquainted enough to choose a field early on. This has given me the possibility to benefit from the most positive aspects of this education system and to be protected from its biases as, not being of French nationality, I was forbidden to enter in the "royal" administrative governmental careers.

I got my first contact with physics there, and thought that this would be useful for an engineering career. At that time there was no job problem especially for students of these schools and the future of my family was totally uncertain after the independence of Morocco. As we had no family fortune, I investigated to join a job in industry (electronic industry?). Being a foreigner I was not proposed the most interesting jobs of the company and I was offered a possibility to work in a department which developed licenses on the radars of the company.

This was a posteriori a real luck, as I found this idea totally unreliable for a 20-year- old non experimented student (I did not know about Einstein!). So I decided that I certainly had still many things to learn before entering an administrative career. I had considered the event of pursuing my education at the University and had passed math examinations at Paris University to open me the possibility to follow a PhD cursus. This was not at that time seen very favourably, and to illustrate that, I just indicate that I got jailed for a week in Polytechnique because I did leave the school courses to pass these math examinations at the University. This was part of the incredible fight between Engineering schools and Universities, which is still vivid in the French community, though no more along the same lines.

One of my Physics professors at Polytechnique (Ionel Solomon) suggested me to meet Jacques Friedel, who had created a modern graduate Physics course on condensed matter in Orsay, where I might be accepted.

## 2.2 First contact with graduate studies

Those preliminaries explain how I have been led to enter J. Friedel's office. As many students of my generation, if not the educational system, nothing would have permitted me to be in touch with J. Friedel who descends from a family with a long university tradition covering more than three generations. He detailed that at will in his essay on the influence of his family on the science policy in France [2].

He has been quite positive about my application, but did not know whether he could get a financial support for a PhD. I did not care, got a small fellowship from the Moroccan government, and began to follow the lectures. In the mean time J. Friedel asked me to apply for an internship position at CNRS, which I obtained essentially because some members of the committee said that there was no reason to make any difference with the French Polytechnique students.

I can tell that, being taught by J. Friedel, A. Blandin, P.G. de Gennes, A. Guinier was definitely my first approach to science. As many students in the French system I had been trained to solve problems which had solutions, kept in a safe, that would be released after the examination. The discovery that we were surrounded by unsolved questions that could be grasped by observation, that the certainties were rather limited, that a professor does not know everything, but qualifies to sort out the actual unsolved questions… was a real shock!

Most of the teaching remained theoretical although, for the first time, a large set of experimental results were quoted in the lectures, showing that these scientists were guided by what comes out of experimental observations. I learned there that strong interplay between experiment and theory was a rule in condensed matter!! During this year of graduate course I also got my first contact with superconductivity and experiments. I had indeed the opportunity to attend, as a master degree student observer to a neutron scattering experiment in which the Abrikosov vortex lattice in superconductors was detected [3]. This was an amazing chance which had an influence on my further choices. Thanks to these experts I did discover that experiment and observations were essential in condensed matter physics, that modelling was the second step and that maths comes only after, in the rationalization process.

## 2.3 PhD and experimental education

The second surprise came when I began discussing about the future possibilities with J. Friedel. He just told me that they had decided to stop the recruitment of theoreticians, the future of the laboratory being to develop experiments. One has to recall that J. Friedel had been working in a small office at Ecole des Mines in Paris and could not get a professor position at Paris University. Thus he had to make the courageous decision to emigrate far from the Latin Quarter, towards a village as far as 25km out of Paris, in association with A. Guinier and J. Castaing to be able to create a laboratory there, a real one with experimentalists! These colleagues were specialized in x-ray and electron diffraction, that were essential to cover the expertise acquired by J. Friedel on point and linear defects, dislocations. But J. Friedel knew that experimental competences on electronic properties had to be implemented there to cover the other part of his theoretical expertise. So J. Friedel took on himself to stimulate an experimental section in the laboratory dedicated to transport properties, magnetism, superconductivity etc… At that time I was not conscious that he was putting together all the elements permitting to build a very modern Material research laboratory.

His proposal for my PhD was then to work in a starting NMR group for which he obtained a CNRS position for Claude Froidevaux who had completed a PhD on low $T$ physics in Oxford and an NMR internship in Berkeley. I am sure today that I could have resisted and insisted to work in theory, but I think I wanted to do something practical, which could be more useful to take a job in industry after PhD. As NMR involved quantum physics in its technical aspects, I did accept, having been attracted towards a PhD by Ionel Solomon who was himself an NMR specialist trained in Abragam's group in Saclay.

I started my PhD then in an empty laboratory, in which a magnet and equipments were due but we were three PhD students, a situation which I have never seen being reproduced in my career- three students, not the worst of the graduate course, with a single advisor, starting simultaneously their PhD. I was impatient and could not wait, so that I decided to accept but asked to spend one year learning the NMR techniques in Solomon's group at Polytechnique, located within Paris at that time. This situation has allowed me to make technical advances in settling the NMR group.

All my PhD was then devoted to an experimental approach on a topic which did not interest much the theoreticians, including J. Friedel himself. This was the Ruderman-Kittel interaction between nuclear spins [4]. These two authors wanted to understand why the NMR widths in pure metals did not fit the expectations from dipole-dipole interactions which did explain very well the situation in insulators. One needed an indirect interaction $J_{12}$ $\mathbf{I}_1$ $\mathbf{I}_2$ between neighbouring nuclear spins $\mathbf{I}_1$ and $\mathbf{I}_2$ through the electron bath that they proposed.

Once understood; this was not such an important matter, but it appeared interesting to try to measure this coupling in different metals. Froidevaux and Weger [5] had discovered that one could measure these interactions using pulsed NMR spectroscopy in alloys. The argument was very simple again. Obviously the NMR were broadened in alloys, due to the Friedel oscillation of the charge densities around impurities, which induced microscopic differences between the Knight shifts of near-neighbour nuclear sites [6-9]. This decoupling of the NMR frequencies of neighbouring sites could be eliminated in a spin echo experiment, so that the remaining exchange interaction $J_{12}$ appeared in the spin echo envelope as a periodic oscillation [5].

So my PhD task had been to improve this methodology and apply it to various possible cases. I was at the end somewhat satisfied of my results [10], and A. Abragam, who carefully controlled my thesis in the PhD committee, appreciated the rigour of the work.

## 2.4 Start of the new laboratory

During that period J. Friedel had enough impact towards the political leaders of the country to obtain the financing of a new laboratory, as we were housed so far in a building loaned by the Department of Physics, in which the possible expansion of the experimental groups was limited. While I developed some pulsed NMR spectroscopy tools which would be important later to develop a group, my PhD advisor got interested in planning the new building and directed its completion until its achievement in 1970. The detailed history concerning the achievement of the "Batiment 510", or LPS, is reported by J. Friedel in his essay on scientific policies [2]. There many students came to follow the condensed matter graduate course for the last 40 years. More importantly, this new building helped to sustain the development of experimental activities and to show the interference between Quantum physics and solid-state physics. J. Friedel was efficiently supported in achieving that aim by the dynamic input of A. Blandin and of P.G. de Gennes who became the fervent supporter of the experimental group on superconductivity.

## 3 From Friedel oscillations to Kondo effect

Before starting to describe the research work I have done with some students, let me recall briefly that C. Froidevaux, immediately after helping to complete the LPS building, left condensed matter physics and initiated a new group dedicated to geophysics within the LPS building. This occurred immediately after my PhD, so that, together with some of my former PhD colleagues, we had the chance to pursue our scientific activity along our wills. As I shall describe below, I felt quite natural to try to work on magnetic impurities in metals, especially with the advent of Kondo effect.

We have seen that the local charge density oscillations associated with a charged impurity in a metal induce a corresponding distribution of the local Knight shifts for the metal host. It was clear as well that spin density oscillations occurred if the impurity is magnetic. Indeed if one considers a local moment on an impurity there appears a difference of phase shifts for the scattering of the spin up and spin down components of spin density. The Ruderman-Kittel interaction that I studied during my PhD was just due to the oscillating spin density induced by the nuclear spin magnetization of a given site of a metal. It is directly felt by the nearest neighbour nuclear spin sites. This approach was easily extended by Yosida [11] to the spin density induced by an electronic local moment (hence the RKKY usually given name).

This had been anticipated as well by J. Friedel and A. Blandin to explain magnetic susceptibilities and the broadening of NMR lines in magnetic alloys. It had been seen early on in Cu-Mn alloys that the copper NMR is broadened without any average shift of the resonance [12]. This was rather easy to understand and only triggered few experimental investigations, until the discovery of the Kondo effect raised important questions. It then became important to study the spin polarization in more details, beyond perturbation theories. Obviously in Orsay the Kondo effect was considered as a first choice theoretical question by J. Friedel, A. Blandin and their students B. Coqblin, P. Lederer, G. Toulouse, etc...

I considered that the questions which arose with the Kondo effect were challenging enough to attempt to investigate some of its experimental manifestations through local NMR measurements. However after five years of my PhD dedicated to experimental efforts I wanted to open my mind by taking a post-doc position for a year just before the completion of the LPS. This was a good chance to take some reflection time before defining a detailed experimental program, which I would start when returning in the new LPS. So I got in contact with Prof. K. Yosida in ISSP Tokyo with a recommendation letter of J. Friedel and was accepted to spend one year in his theory group.

This was an interesting situation, as I was not a theorist, having never been trained to perform theoretical calculations with Feynman's diagrams, contrary to all the Japanese post-docs in Yosida's group. So I concentrated my efforts on understanding the open theoretical problems on Kondo effect and specifically about the questions which could be answered by taking NMR data. I could as well perform some basic calculations which could be helpful for the NMR data analysis. This has resulted in a series of experiments in the years 1971-78 that I shall sketch hereafter without sticking on the actual chronology. I shall rather try to reproduce the state of the art which was achieved and remains classic. I shall first describe how we could monitor the spin density oscillations accurately in § 3.1 and see how they evolve with $T$ in the case of a Kondo impurity in §3.2. There, I shall show as well how I could carefully study the dynamic properties of the impurity electron spin, and get altogether simple evidences about the low $T$ singlet formation and dynamics. I shall briefly consider in §3.3 the extensions of this work which lead to Heavy Fermions physics and the questions that were and are still pending on the Kondo problem, as they have been recently underlined again by theorists like I. Affleck. The conditions which have permitted the positive results obtained during these years will then allow me in §3.4 to discuss the importance of the dialog between theory and experiment, and how the material research aspects could be handled successfully in the structure which had been put together by J. Friedel at the LPS.

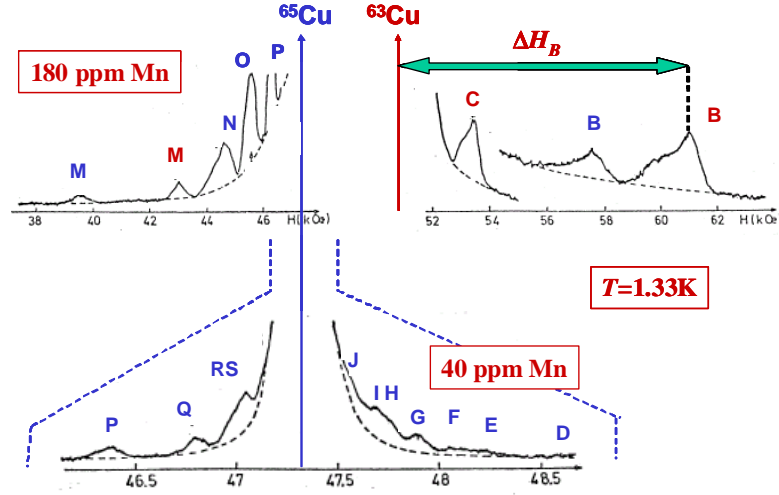

**Fig. 1** NMR spectra of the $^{63}$Cu and $^{65}$Cu nuclear spins in dilute Cu-Mn alloys obtained by sweeping the applied external field, at 1.3K. The spectra have been expanded vertically to exhibit the satellite lines by cutting the large intensity central lines at the pure copper NMR positions, which are pointed by arrows. The near-neighbour satellite lines displayed there are labelled as A,B,….on the right and M,N,O… on the left. Here the A line is out of the field scan and has been detected at higher temperature. The sites far from the impurities are better resolved by reducing the impurity concentration, as shown in the expanded bottom spectrum (adapted from Ref. [13]).

### 3.1 Magnetic impurities and spin density oscillations

For local moment impurities, Yosida [11] calculated explicitly the spin density oscillations assuming that the free electron spin **s** and local moment **S** interact by an exchange interaction

$$H = -J\ \mathbf{S}.\mathbf{s}\,\delta(\mathbf{r}) \ . \qquad (1)$$

The resulting local spin density, calculated in perturbation theory at a position $R_n$ with respect to the impurity is given by

$$n(R_n) = -\frac{1}{4\pi} J\rho(E_F) \frac{\cos(2k_F R_n)}{R_n^3} <S_z>, \qquad (2)$$

for a field applied in the $z$ direction. So, for local moment impurities such as Mn in Cu the NMR shift of a Cu nuclear spin at position $R_n$ with respect to the impurity acquires an extra NMR frequency shift $\Delta K$ given in an applied field $H$ by

$$H\Delta K = A_{hf}\ n(R_n) = A(R_n) <S_z> \qquad (3)$$

where $A_{hf}$ is the on site Cu hyperfine coupling, and $A(R_n)$ is then a transferred hyperfine coupling between the local moment and the nuclear spin at $R_n$. Using Eq (2), one can see immediately that this extra shift oscillates around zero, so that in an alloy the main effect is a distribution of shifts which is mainly reflected as a broadening of the Cu NMR without any mean displacement of the line. This had been seen quite early on alloys with a sizeable impurity concentration [12]. But such restricted information only permitted to control that the $T$ and $H$ dependence of the induced host NMR width reproduces the Brillouin-like variation of the impurity magnetization.

The occurrence of the Kondo problem raised questions about the applicability of this simple relationship, which triggered more detailed experimental studies. However, as long as one had only the overall broadening of the NMR as experimental information, this did not allow any efficient study of $n$(r). But progressive technical improvements, and the use of high magnetic fields allowed by the advent of superconducting magnets, permitted to resolve the spectra and measure the spin polarization on different shells of near-neighbour sites of the impurities. Though this was initiated in the important case of Fe in copper, which will be discussed in the Kondo effect section, I shall report first here the remarkable results that we obtained later in Cu-Mn.

#### 3.1.1 NMR detection of the spin density oscillations

In most dilute alloys of transition elements only a few near-neighbour shells of the impurity could be resolved. However in Cu-Mn, the impurity magnetization which follows a Brillouin function can become quite large at low $T$ and high fields, so that the magnitude of $n(r)$ can become large enough on many neighbouring sites of the impurity. This permitted us to resolve up to 17 distinct shells of neighbours, as can be seen in the spectra of Fig1, which gives a straightforward illustration of the occurrence of spin density oscillations [13]. Indeed, one can see that there are about as many extra lines (we call them satellite lines) on the right and on the left of the central line. The simple equation (2) has spherical symmetry around the impurity so that each satellite is expected to exhibit a similar width as the central copper resonance, which is governed by the distribution of fields due to distant impurities. In reality, we can see in Fig 1 that some of these satellites, such as B, have broad non-symmetrical lineshapes, which could be associated with an angular-dependent spin polarization. This indicates that the exchange coupling of conduction electrons of (2) has to be complemented by taking into account non-axial dipolar couplings [14]. Using site symmetry arguments, this did help to identify the sites corresponding to these lines [15], and we could as well assign some of the other satellites, by comparing the relative intensities with site occupancies.

This analysis allowed us to reproduce the experimental shape of the spin density oscillations as given in Fig. 2, though all the satellite lines detected could not be assigned unambiguously. There it can be seen that the data agree with the overall $r^{-3}$ dependence, but do not follow the asymptotic limit at short distance as will be discussed in the next paragraph.

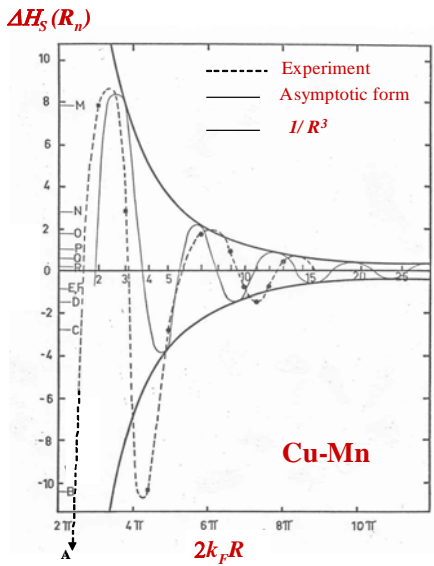

**Fig. 2** The value of the field shift at saturation of the magnetization for the most shifted satellite lines of Fig. 1 is plotted versus the assigned position $R_n$ of the corresponding sites (the horizontal scales report the value of the shell number $n$ in the middle axis and that of $2k_FR_n$ on the lower axis). Notice that the satellites corresponding to the most distant sites, that is the smallest shifts in Fig. 1 are only tentatively assigned. The comparison of the data with the asymptotic function given by Eq. (2) is discussed in the text (from Ref. [13]).

### 3.1.2 RKKY and Friedel-Anderson model

While the data on the satellites only permits to study the short-distance behaviour of $n(r)$, the use of the information embedded in the central lineshape and width is quite important to obtain the magnitude of the long-distance behaviour of $n(r)$. The numerical simulations I had performed initially [16], when no NMR satellite data were available, had allowed us to compute the expected variation of the half intensity width $\Delta H_{1/2}$ with impurity concentration $c$, and to verify that the $r^{-3}$ asymptotic behaviour of Eq. (2) naturally yields a computed linear $c$ dependence given by

$$\Delta H_{1/2} = \alpha \, J\rho(E_F) <S_z> c , \qquad (4)$$

where $\alpha$ is a coefficient deduced from the computation. This asymptotic limit holds as long as the impurity concentration does not exceed 1%. The validity of this expectation was confirmed experimentally and permitted to deduce $J\rho(E_F)$. By combining Eqs. (2) and (3) we obtained the radial dependence of $n(r)$ plotted as a thin full line in Fig. 2. There it can be seen that the magnitude and the oscillatory behaviour of $n(r)$ applies perfectly down to distances corresponding to a few shells of neighbours. However the phase of the oscillatory behaviour deviates markedly from Eq.(2).

Such a pre-asymptotic deviation with respect to (2) was indeed expected, as the exchange interaction of (1) only takes into account the spin properties of the impurity, and does not consider its detailed electronic structure. It was clear from discussions with A. Blandin and J. Friedel that in the more realistic Friedel-Anderson approach one needs to consider the scattering of the conduction electrons by the impurity level, which in a Hartree-Fock approximation is represented by a nearly filled virtual bound state (VBS) for the case of a magnetic impurity. In this approach the Coulomb interaction $U$, the position and width of the $3d$ level should be reflected in the spatial dependence of the spin polarization. While in Tokyo, I had done such calculations under the guidance of H. Ishhi and K. Yosida, and did indeed show [16] that, as displayed in Fig. 3, the preasymptotic behaviour of the spin polarization was quite different for the exchange $sd$ model with $d$ wave scattering and for the magnetic Anderson model.

This incidence of the electronic structure of the impurity level on the spin polarization shape could only be studied through experiments on the satellite NMR lines. This has been completed in the thorough work done in C. Slichter's group. There, various students did investigate the whole $3d$ series of transition elements in copper [17]. They could show that the polarization on the few first shells of neighbours is quite dependent on the electronic structure of the impurity levels, and this could be reproduced later quantitatively by full LDA calculations [18].

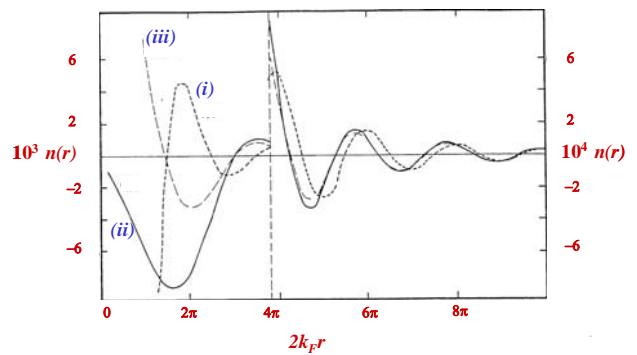

**Fig. 3** Results of calculations of the spin density oscillations in three cases (i) asymptotic RKKY limit of Eq. (2), (ii) d wave exchange and (iii) magnetic symmetric Anderson model. Here one can see that the most prominent modifications occur for $2k_Fr<4\pi$, that is on about four shells of neighbours of the impurity (from Ref. [16]).

### 3.1.3 Non magnetic limit

Another case which was important to study as well was that of weakly magnetic impurities which do not display local moment behaviour but rather correspond to the non magnetic limit of the Anderson model in the Hartree-Fock sense, for which an unsplit VBS occurs at the Fermi level. The best known situation of that kind was initially that of Al-Mn for which the impurity spin susceptibility displayed only weak $T$ variation [19]. Indeed, as can be seen in Fig. 4 the calculations [16] confirmed that one expects a large preasymptotic reduction in the magnitude of the spin polarization for distances such that $k_F r < k_F r_c = \Delta/E_F$, where $\Delta$ is the width of the VBS resonant state at the Fermi level. We computed that this should be reflected in the concentration dependence of the NMR width, which is expected to decrease for large impurity concentrations with respect to the $c$ linear dilute limit of Eq. (4).

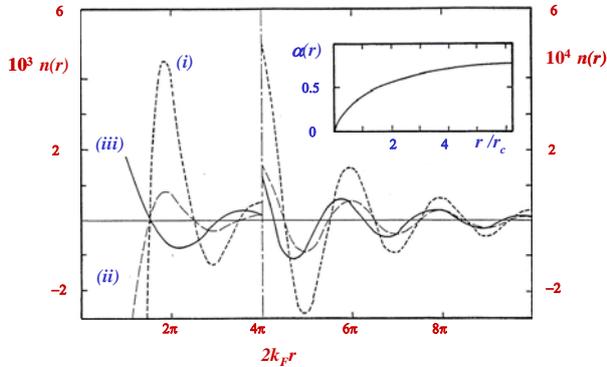

**Fig. 4** The spin density oscillations **(i)** for the asymptotic limit of Eq. (2) is compared with that for the non-magnetic Anderson model with a VBS at the Fermi level of width $\Delta = E_F/10$ (**(ii)** is a computation result, while **(iii)** represents an approximate analytic expression). The relative decrease in magnitude $\alpha(r)$ of the oscillations for **(iii)** with respect to the $1/r^3$ asymptotic limit is plotted versus $r/r_c$ in the inset (from Ref. [16]).

Experimentally, the magnetic effects being somewhat small in this case, the width of the $^{27}$Al NMR had only a very weak $T$ dependence. The magnetic contribution to the broadening was rather small as well, and its determination was limited by the pure aluminium metal width due to dipole-dipole interactions between $^{27}$Al nuclear spins. We could however reduce this contribution to the width, using specific multi-pulse experiments or high applied fields [20]. The magnetic width deduced that way definitely departs from linear behaviour versus concentration above about c= 500ppm, so that the asymptotic limit might only apply at very large distances from the impurity. We also detected in Al-Mn two satellite resonances [19] on the same side of the main line, which are much less shifted than expected from the asymptotic limit. So, overall these results do establish that, contrary to the case of Cu-Mn, the magnitude of the spin polarization is depressed near the impurity with respect to the simple asymptotic $r^{-3}$ variation. Assuming that the non-magnetic solution of the Anderson model applies, a fit of the data could be achieved assuming $\Delta = E_F/10$.

So all the experiments quoted above indicated that the electronic structure of the impurity level is reflected in the spatial dependence of the induced spin polarization. This opens then quite naturally important questions concerning the evolution of this spin polarization in the Kondo state of a magnetic impurity that we shall consider next.

### 3.2 The Kondo problem

The Kondo problem arose with the discovery by Kondo [21] that perturbation theory of the *sd* Hamiltonian of Eq. (1) resulted in a $-\ln T$ term in the resistivity of the alloys, which was indeed observed experimentally. It was understood that the conduction electron interaction with the local moment induced a crossover of the electronic state towards a low $T$ ground state quite different from the quasi-free local moment and that the crossover temperature defines an energy scale

$$k_B T_K = E_F \exp\left[1/J\rho(E_F)\right] \qquad (5)$$

This expression for the Kondo temperature $T_K$ bears some analogy with that of $T_c$ and the energy gap variation with electron-phonon coupling for superconductivity. It has been harder to qualify the actual properties of the Kondo ground state, but from the observed transport and thermodynamic properties associated with the impurity degrees of freedom, it has been accepted rather soon that the impurity properties experimentally appear to evolve from a high $T$ magnetic state to a non-magnetic like behaviour below $T_K$. In other words, the weak coupling regime where the impurity moment can be treated in a perturbation scheme evolves at low $T$ in a strong coupling regime where the impurity and conduction electrons are bound into the ground state. The basic picture which was initially accepted is that the conduction electrons might form a singlet state with the impurity and compensate its magnetization. If such a spatially extended state occurs, one would then expect to see its experimental signature on local magnetic measurements in the corresponding spatial range around the impurity, so that NMR experiments would be the ideal probe to view such effects.

From the study of the macroscopic properties of impurities in noble metal hosts, it was established that the crossover temperature $T_K$ was highly dependent on the impurity [22]. This was of course quite compatible with the exponential expression of Eq. (5). $T_K$ could be estimated from the maximum in the impurity contribution to the specific heat, or from the Weiss contribution to the spin susceptibility measured at high enough temperature, etc... This permitted to establish that $T_K$ was below 10mK for Cu-Mn, ~ 1K for Cu-Cr, ~ 30K for Cu-Fe, ~ 300K for Au-V, etc....It was harder to consider Al-Mn along the same lines as all temperature dependences were very weak in this case, so that this crossover could only occur above 1000K, for which the alloy would have molten.

Anyway, if one wants to study the change from the magnetic state to the non magnetic state, one needs to consider in priority a system in which one can explore both regimes $T \gg T_K$ and $T \ll T_K$.

So Cu-Fe appears immediately as the most suitable case if one wants to avoid extremely low temperature experiments, while Cu-Mn and Al-Mn appeared as the two extreme cases.

### 3.2.1 Spatial extent of the Kondo singlet?

This idea of a Kondo singlet emerged at a time where the various satellite lines could not be resolved, so attempts were done to detect modifications of the host NMR width when $T$ is decreased through $T_K$. Measurements done in the group of A. Heeger (chemistry Nobel prize in 2000) revealed that the $^{63}$Cu width increased below 30 K somewhat faster than expected from the high $T$ Curie-Weiss susceptibility, and saturated at low $T$ [23]. This was initially taken as a signature of the development of a static polarized cloud anti-parallel to the local impurity magnetization. These data raised some doubts as there were indications that the macroscopic magnetization of Cu-Fe alloys did not grow linearly with impurity concentration, suggesting that interactions between impurities were at play [24]. Our own data on the NMR width in more dilute samples did not display either such a large increase in width below $T_K$ [25].

The situation was only fully clarified when satellite NMR resonances similar to those presented in Fig. 1 were detected both by Slichter's group [26] and myself [27]. We immediately found that the shifts of the various satellites had $T$ dependences which scaled with each-other and displayed the same Curie-Weiss variation as the magnetic susceptibility data taken in very dilute samples, as displayed in Fig 5. So, on a small number of sites near the impurity, the magnetic behaviour did display a smooth $T$ variation through $T_K$, which allowed us to deny the existence of a static compensating cloud.

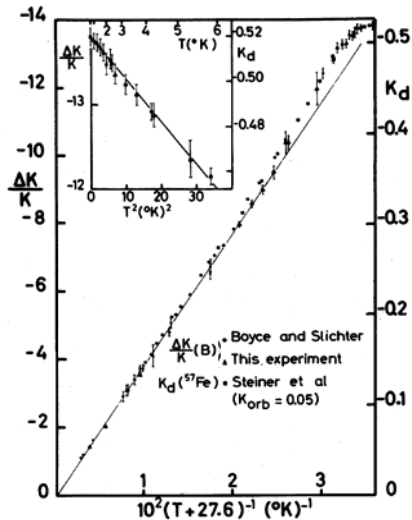

**Fig. 5** The variations of the normalized NMR shift $\Delta K/K$ induced by Fe impurities in Cu for three satellite resonances and of the impurity susceptibility obtained by Mössbauer effect data ($K_d$) scale perfectly with each-other. This gives a good experimental determination of the variation of the local spin susceptibility through $T_K \sim 30$ K, which crosses over from a high $T$ Curie-Weiss dependence towards a quadratic $T$ variation below $T_K$ (see inset) (from Ref. [27]).

Let us point out that the results which I display in Fig. 5 permitted us to:
- Demonstrate that the local susceptibility measured on the satellite NMR lines scaled perfectly as well with the local susceptibility on the Fe sites, as given by Mössbauer effect data, and is not modified below $T_K$ when the Kondo state is established.
- Provide the best determination of the $T$ dependence of the local moment susceptibility which crosses over from a Curie-Weiss law above $T_K$ to a $T^2$ dependence below $T_K$ [28]. This result confirms that the susceptibility reaches a low $T$ behaviour similar to that achieved in a non-magnetic case, as has been also found by the numerical solutions of the Kondo model established [29] by Wilson (Nobel prize) at about the same time.

Furthermore the positions of the satellite lines were independent of Fe concentration, as the nuclear spins sense the properties of isolated impurities. Indeed, as was anticipated by the analysis of the low $T$ susceptibility data, the spin polarization around groups of Fe impurities strongly coupled to each-other are not expected to display the same $T$ variation as isolated impurities, so the NMR signals of the corresponding Cu sites are weak and buried in the smooth background NMR signal. However, these results do not give any answer about the spatial extension of the correlated Kondo state, a matter which I shall discuss in § 2.3 after considering the impurity dynamic susceptibility.

### 3.2.2 The dynamic response

A large effort has been done initially using ESR spectroscopy to try to understand the dynamic properties of local moment impurities in metals. It is well known that nuclear spins **I**, which couple to the band $s$ electron spins in a metal by a scalar contact hyperfine coupling

$$H = A_{IS}\, \mathbf{I}.s\, \delta(\mathbf{r}), \qquad (6)$$

display a spin lattice relaxation rate $(T_1)^{-1}$ given by the Korringa expression

$$(T_1)^{-1} = (\pi/\hbar)\, k_B T\, A_{IS}^2\, \rho^2(E_F). \qquad (7)$$

The local moment **S** being coupled to the band electron spins by the analogous Hamiltonian of Eq.(1) one then expects to obtain a similar local moment relaxation rate

$$\tau^{-1} = (\pi/\hbar)\, k_B T\, J^2\, \rho^2(E_F). \qquad (8)$$

In principle, this relaxation rate should be seen as a broadening of the local moment ESR. However it had been demonstrated for long [30], for instance in the case of Cu-Mn dilute alloys, that the coupling **s.S** is so large that one detects a single EPR signal which results from the combined dynamic response of the magnetizations of the two sets of electron spins. This so-called bottleneck effect prohibits the determination of $1/\tau$. Furthermore, if the local moment is spectrally decoupled from the conduction electrons, for instance by a large difference of g factors, the impurity EPR signal usually becomes so broad that it cannot be detected for 3d impurities in metals.

So we had been attempting to use the nuclear spins as probes of the dynamics of the local moment. Indeed the occurrence of the spin density oscillations around the impurities, recalled in the previous sections resumes in an indirect coupling between the nuclear spin at site $R_n$ and the local moment at the origin, given by

$$H_{IS} = A(R_n) \; I_n \cdot S \; . \quad (9)$$

It can be easily seen that the fluctuations of the local moment induce local field fluctuations on the nuclear spin $I_n$ at site $R_n$ that are responsible for its nuclear spin lattice relaxation rate. The latter can be written as

$$(1/T_1)_n = 2\hbar^{-2} k_B T \; A^2(R_n) \frac{\chi}{(\hbar\gamma_e)^2} \frac{\tau}{1+(\omega_e\tau)^2} \; . \quad (10)$$

Here one assumed a standard Lorentzian shape for the local moment dynamic susceptibility. As the coupling $A(R_n)$ decays as $1/R_n^3$, the nuclear spins displaying the highest relaxation rate are the nearest neighbours of the impurity. Those contributing to the host main line NMR are weakly coupled to the local moment and display a distribution of spin lattice relaxation rates. We could show that beyond a certain distance, the local $(T_1)_n^{-1}$ becomes small, so that the nuclear spin magnetization diffuses fast enough to display a uniform $(T_1)^{-1}$ decay for long times [31]. The full analysis of this process allowed us to conclude that the fluctuation rate $1/\tau$ of the local moment does indeed vary linearly with $T$ for Cu-Mn, [32] and deviates from this variation when crossing the Kondo temperature in Cu-Fe.

However, more quantitative data could be obtained when it became possible to detect the NMR signals of the individual near-neighbour shells of the impurity. We could then combine the $T_1$ data with the measure of $\Delta K(R_n)$, the contribution of the local moment to the shift of given satellites This eliminates $A(R_n)$ between Eqs. (3) and (10), and permits to deduce $1/\tau$. The data for this quantity shown in Fig. 6 are indeed found independent of the observed satellite, and are furthermore $T$ independent below $T_K$ in Cu-Fe.

In the low $T$ limit, the data resumes into a Korringa relation [28]

$$T_1 T \Delta K^2 = \frac{\hbar}{4\pi k_B} \left(\frac{\gamma_e}{\gamma_n}\right)^2, \quad (11)$$

where $\gamma_e$ and $\gamma_n$ are the electron and nuclear gyromagnetic ratios.

We therefore established that $\tau^{-1}$ reached a constant value in the Kondo state, as does the local moment susceptibility. This also means that the Kondo energy scale $k_B T_K$ which limits the increase of $\chi$ below $T_K$, governs as well the $T$- independent spin lattice rate $\tau^{-1}$ for the local moment, with both $\chi \sim \mu_B^2/k_B T_K$ and $\hbar/\tau \sim k_B T_K$ [33]. Such a result then fits with the idea that the Kondo state has a dynamic behaviour controlled by $k_B T_K$.

Let us point out that calculations done along the RGA or Bethe Ansatz showed that the Kondo state displays a resonance at the Fermi energy [34], which has a very narrow width controlled by $k_B T_K$, and that the local moment spin lifetime appears as a direct determination of the width of this Kondo resonance peak. We could as well do similar measurements of $1/\tau$ in Cu-Mn [14] which could be combined with the data taken in Cu-Fe to give the overall variation of $1/\tau$ over a few decades in $T/T_K$ extending below and above $T_K$.

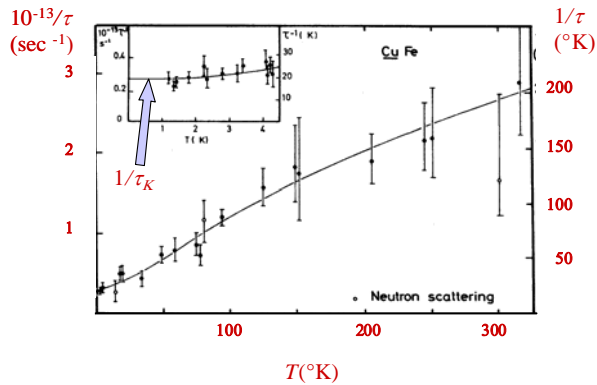

**Fig. 6** Experimental variation of $\tau^{-1}$ deduced from the $(T_1 T)^{-1}$ data for various near-neighbour sites of Fe in copper. Three data points obtained from neutron scattering experiments are plotted with empty symbols. The inset gives an expansion for the low $T$ scale values (adapted from Ref. [27]).

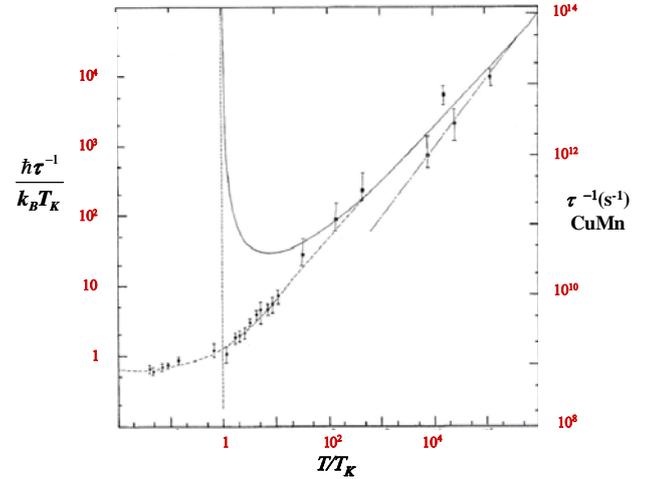

**Fig. 7** Experimental variation of $\tau^{-1}$ (right scale) deduced from the $(T_1 T)^{-1}$ data for various near- neighbour sites of Mn in Cu ($T_K$=0.01 K, filled squares). The data for $\hbar \tau^{-1}/k_B T_K$ (left scale) are compared with those of Fig. 6 obtained in Cu-Fe ($T_K$=30 K, filled circles). All scales are logarithmic. The full line corresponds to the result of first- order perturbation theory, the dash-dotted line is the Korringa limit, and the dashed line is a guide to the eye interpolating between CuFe and CuMn data (from Ref. [14]).

### 3.2.3 Recent developments

Very extensive studies of Kondo rare earth impurities in metallic matrices have been done as well. In most cases it is again quite difficult to investigate large $T$ scales above and below $T_K$, which are usually rather low. Those could be measured in Orsay by our colleague J. Flouquet, using original Nuclear Orientation experiments

[35], before he left for Grenoble. Of course the most interesting physical effects still vividly studied nowadays are those which appear in dense Kondo lattices, as they give rise to most of the original properties of the metallic rare earths studied at length by Coqblin [36] and to Heavy Fermions physics [37]. The novel development of ARPES techniques has permitted to evidence as well the Kondo resonance in the energy response function in Kondo lattices, as one can benefit of the wave vector sensitivity of this technique [38].

A large extension of the Kondo physics has also been seen in mesoscopic physics, for which a quantum box with a small number of electrons behaves as an artificial atom interacting with incoming metallic leads [39].

Finally the development of new experimental local probe techniques such as STM has recently permitted direct observations of Friedel oscillations. Just after the Nobel price awarded in 1986 to Binnig and Rohrer for the discovery of this technique, low $T$ STM experiments at IBM Almaden, gave evidence for local charge density oscillations induced by step edges or point defects on the surface of Cu metal [40]. This has been regarded as one of the most beautiful evidence of Friedel oscillations as they permit nice 2D images. However, I still feel that few experiments can so far give as nice quantitative determinations of the oscillating spin polarization induced in 3D systems by magnetic impurities as that reported here and displayed in Fig. 1.

Concerning the most classical Kondo physics, most theoretical developments on Kondo impurities have been addressing the energy dependences which are responsible for the variations of the thermodynamic and transport properties. These results matched rather well the experimental observations, although it is not yet clear whether multi-orbital cases have been treated sufficiently well to permit detailed quantitative comparisons.

Further, at the time of our experimental work, the theoretical calculations on Kondo effect did take detailed considerations of the actual spatial signatures of the Kondo length scale. We were of course somewhat surprised to find out that the spin polarization near the impurities follows exactly the magnetic data on the impurity. I however discussed that at length with H. Ishii, a former post-doc of K. Yosida who visited Orsay for more than a year, and he did theoretical calculations in agreement with this observation [41]. The spatial extent $r_c$ of the Kondo correlated state is easily estimated as being given by $k_F r_c = E_F/k_B T_K$. Our results have shown that we could not detect in Cu-Fe any change through $T_K \sim 30K$ which could be associated to the formation of the correlated singlet state. But it was easy to see that $r_c$ is so large that one needs a huge experimental resolution to detect anything connected to this Kondo length-scale. Recently I. Affleck has again addressed that point theoretically and underlined that difficulty [42-43], and calculated the expected variation through $r_c$. The only case where something would be detectable is that of systems with extremely high $T_K$, which could be that of the "non-magnetic" Al-Mn that we studied in §3.1.3. This case would correspond to a Kondo state for which $T_K$ would be as large as 1000K. Then, the Kondo length-scale $r_c$ should be much reduced, by a factor 30 with respect to that of Cu-Fe. The attempts we did to investigate the shape of the spin polarization in that case revealed that the asymptotic limit does not apply to short distance from the impurity, so there we might indeed have been sorting out a spatial variation of the spin polarization associated with the Kondo scale. Though in this system $T$ dependences cannot be investigated, as data can only be taken for $T < T_K$, further refined studies should benefit of the large improvement in NMR techniques done over the last 30 years.

### 3.3 Some comments about the achievement of these experiments at LPS

All the experiments depicted above have been done in the LPS in the few years which followed our instalment in this new laboratory. When looking back at that period, this was in my opinion a golden age of science. This was a fantastic period where physical understanding was the main objective. We did not care at that time about career development, nobody asked us to put any emphasis on potential applications of our results. We were essentially driven by intellectual understanding. Our level of international recognition, which was initially at the ground level, raised progressively when we sharpened our experimental tools and obtained significant results.

I would like to insist here that these successes did not result only from personal factors of the researchers involved. Obviously some alchemy had been performed by our advisors when they decided to group altogether not only theorists and experimentalists on condensed matter, but also material research groups. For the work presented here–above, many dilute alloys had to be synthesized. We largely took benefit from the existence of a metallurgy group which has certainly been driven here as a consequence of the large importance J. Friedel has always given to material studies. That did not mean that we ordered our samples to that group, but we learned there how to make them ourselves in the various furnaces and equipments which were implemented in the laboratory. That was not only a bonus in our education, but as well a great chance to become aware that we were working on materials which are not as ideal as one would like them to be. Intrinsic solubility limitations and inhomogeneous impurity distributions did occur in these alloys. Therefore significant material research work had to be done to get reliable samples and to optimize them specifically for the experiments we wanted to perform.

In one sense, as I have already stressed above, the education we got in LPS allowed us to be at ease both with theoreticians and with material science aspects. This was not at all in our undergraduate education, but was acquired totally during the achievement of our experimental activities. The structure of the LPS was such that all this was on free access, and that most of my colleagues were picking the points which they felt best suited to their scientific aims. To give a simple image, I would say that our seniors, J. Friedel, A. Blandin, P.G. de Gennes, A. Guinier and R. Castaing had put together a "bistrot" in which many nice bottles were stored including theoreticians doing analytical calculations, others using computers intensively, various competitive experimentalists with specific advanced techniques, plus some materials research groups in metallurgy, organic chemistry etc... We were then somewhat free to concoct the cocktail most adapted to our taste. I am convinced that many of my colleagues, including I. Campbell, A. Fert, J. Flouquet, D. Jérome, H. Launois, P. Monod etc..., would not disagree with this description of the situation which has prevailed during the 1970's, and which has been important as well in their scientific achievements.

Concerning the theoretical aspects on Kondo effect, I had been discussing quite often with A. Blandin and J. Friedel and with my Japanese colleagues as I explained above. It was clear in that period of time that theory and experiments were highly connected with each -other and that both were advancing in parallel. In about 10 years' time a consensus emerged at the international level and an understanding of the correlated Kondo state was established.

I think that the situation I depicted above has to be regarded as quite original, especially in the French context, as the weight of the "Descartes classification" is so heavy here. That quite naturally has led to some sociological problems magnified by the fact that many motivated young scientists had been recruited altogether in a short-while. I prefer to avoid entering into problems which only deal with the French microcosm, and summarize positively here that this has permitted us to express our skills independently and to enter the international competition.

At the personal level, I could train during that period some students in directions inspired by the overall scientific activity at LPS. Though I have been working on various topics in the 1980's, including spin glasses and insulating glasses, I kept a strong interest in the correlated electron physics which I touched for the first time in these studies on Kondo effect. I did then some work on nearly ferromagnetic metals [44] and on the insulator to metal transition in phosphorus -doped silicon [45], where correlation effects were found more important than expected. During that period, most of the remarks on the scientific context which I have depicted above remained valid.

## 4 The cuprate wilderness in correlated electron physics

The discovery of High $T_c$ cuprates in 1986 has been a fascinating event for most researchers in condensed matter physics. Let me recall that the "Woodstock" of cuprate physics occurred at the APS March meeting in 1987. The occurrence of the high $T_c$ values, exceeding liquid nitrogen temperatures, has led many researchers to be imprudent enough to claim that room temperature superconductivity would be accessible in a short-while. But on the scientific side it was clear for many others that the most interesting point was that these high $T_c$ values occurred in materials which were doped Mott insulators. This has led many researchers to infer that this superconductivity was unconventional.

These astounding interests have led in the last 20 years to superb developments of new techniques but also to confusing interpretations and speculations. There would be too much to say here on many of those aspects. As the scientific issues are still highly debated at the time I am writing this contribution, and as any scientific discussion of recent results would need extensive developments in which J. Friedel has not been involved, I shall try here to restrict mostly to a presentation of the initial work done at LPS.

First, I shall explain in §4.1 how the internal scientific organization of the LPS permitted us to highlight the importance of materials aspects and to begin to master some of them. I shall then describe in §4.2 the approach which has led my group to produce significant contributions in the early days of the cuprates. Those were performed while J. Friedel was still very active at the LPS. A few comments on work on the "pseudogap" physics and the phase diagram of cuprates in which I was involved later with collaborators of my group or of the Saclay area will be given in § 4.3

### 4.1 The "prehistoric" investigations of HTSC at LPS

At the beginning of 1987 all the researchers in the LPS attempted to work on High $T_c$ cuprates that is on the $La_2CuO_4$ family discovered by B. Raveau in Caen and highlighted as being HTSC by Bednorz and Muller. Though, as usual, some were trying to govern the scientific activity by trusting samples, the fact that we had in the laboratory and the university some researchers quite acquainted with material research prevented such negative attempts.

For my group, the situation was initially extremely simple, as we were nearly totally unable to do any work. Indeed, our superconducting NMR magnet, which was one of the preliminary versions manufactured early on in 1965, was damaged in 1981 and could not be repaired. This was the magnet in which we had done most of the experiments underlined above on Kondo effect. By chance I was able to start a program on zero-field NMR experiments in the frozen state of spin glasses [46], in which P. Mendels was involved for his PhD. It took overall four years to collect the financial support required to order a magnet with a room temperature bore, in which we planned to house a $He^3$- $He^4$ dilution cryostat for some experiments on phosphorus-doped Si. We also intended to use a $He^4$ cryostat to work on superconducting Chevrel phases, but had not even anticipated to take data between room and $He^4$ temperatures. Furthermore, technical difficulties encountered by the manufacturer delayed the delivery until September 1987. So we were technically unable to enter the scientific competition on the 124 compounds.

But of course experimental investigations were possible in most other research teams, and many working groups and seminars in the Orsay area permitted exchanges with theoreticians and allowed us to follow the fast evolving situation at the international level. Most of the initial guesses and speculations about the HTSC phenomenon have been started on the basis of the results found in 124 compounds. The doping being achieved by substitution of La by Ba or Sr, it was immediately found that the AF state can be destroyed easily for ~2% substitution, and that the ground state is a spin- glass non-metallic phase up to 5% substitution, above which metallic behaviour and SC appear.

This has naturally led many researchers, including J. Friedel, to consider that magnetism competes with SC. Furthermore large single crystals of these compounds could be produced in a short-while, and inelastic neutron scattering peaks of magnetic origin were found to occur at incommensurate wave vectors which changed continuously with doping. It was then somewhat natural to consider that the magnetic dynamic properties in the doped regime could be described in a progressively filled band in which correlations decrease upon doping. As the maximum of $T_c$ occurs there, these results had led J. Friedel to anticipate that the AF correlations are not important in the HTSC phenomenon and to oppose the trend favoured by the tenants of strong correlations who considered that the important starting point was that the parent compound is a Mott insulator.

It was then quite justified to extend the approach which successfully explained the rather high $T_c$ of the A15 metallic compounds like $V_3Si$. J. Friedel proposed that those observed in the cuprates could be due to the enhanced density of states associated with the Van Hove singularities of the 2D electronic structure. This has been clearly spelled out in an article [47] that J. Friedel wrote later for the M2S-HTSC conference in Interlaken in 1988, which quotes most work done at the LPS. Various contributors to this Festschrift have shared similar ideas and will certainly relate this point in a detailed manner.

I shall rather consider here the incidence of the discovery of $YBa_2Cu_3O_{6+x}$ by Chu and collaborators in Houston, which was reported at the "Woodstock of Physics". With $T_c$ values as high as 90K, that is above liquid nitrogen temperature, this material reinforced largely the general interest and later opened possibilities to demonstrate superconductivity and Meissner effect to students first [48], and quite beautifully recently to the general public [49]. In Orsay, we had in the laboratory and the university enough personalities able to synthesize these materials, and ready to provide samples and to collaborate with anybody willing to work on them. The new experimental windows opened by the discovery of this compound have emulated a stimulating competition between groups inasmuch as new theoretical questions arose.

Interestingly I did visit in spring 1987 some laboratories in Japan within a French-Japanese exchange program planned for long. The situation was extremely competitive as well there and I remember discovering on that occasion what I shall call the "HTSC wilderness", as I was not allowed to enter Prof. Yosuoka's laboratory at ISSP but only to discuss with the students on any work except cuprates. They were clearly not ready to disclose any "secret" as I was obviously considered as a potential spy. How different was the situation with that I had known during my post-doctoral internship!

During that period, as I explained above, the only possible attitude for me was to study the huge flow of arriving publications through the "High $T_c$ update" abstracts archives channel, which was transformed later into the condmat archives sections. This is where I discovered that all NMR specialists were working quite naturally on $^{63}Cu$ NMR to get information on the SC properties. But I noticed a single work on $^{89}Y$ NMR, which revealed that, though out of the planes, this nucleus was still weakly coupled to the electronic properties of the planes and could be used to monitor them [50]. I immediately considered that this opened a possibility to study by NMR the magnetic responses of weakly metallic phases.

In any case I knew that the fixed-field homogeneous magnet would not be adapted to study the broad $^{63}Cu$ NMR signals which were detected in the cuprates. So I kept the idea to study the $^{89}Y$ NMR and looked at the possibilities to control the doping of these materials in order to be able to test the magnet with such experiments. The contribution of G. Collin at that period has been quite essential, as he provided us optimally doped powder samples that he checked carefully by x-rays, and this has been the start of a long-lasting collaboration with him on cuprates and more recently on layered cobaltates.

With P. Monod we also understood that it was rather easy to change the oxygen content of these materials and to get samples with distinct dopings. We had put together a furnace in which we could anneal $YBa_2Cu_3O_7$ samples at different temperatures and used initially the weight loss to estimate the actual oxygen content. Later on, we collaborated with J.F. Marucco from the Solid State chemistry laboratory in Orsay, who could do that in situ in a thermo-balance, which permitted him to achieve more accurate thermo-gravimetric control. This interaction with groups controlling material research aspects became essential to develop experiments versus oxygen content on samples with impurity substitutions.

As soon as it was established that the parent $YBa_2Cu_3O_6$ compound was AF in its ground state, as was $La_2CuO_4$, it became clear that it would be essential to understand the evolution of magnetic properties with doping, so that we began to test macroscopic properties for varying oxygen content and did not focus on the higher $T_c$ compositions.

On the NMR side, the equipment designed with P. Mendels for the zero-field NMR studies on spin glasses could be used to try to discover the zero field $^{63}Cu$ NMR signal in the AF state of $YBa_2Cu_3O_6$ [51]. Meanwhile, I was progressively preparing the equipment which could allow us to perform the $^{89}Y$ NMR experiments upon delivery of the magnet. Both approaches happened to be successful, the first one allowed P. Mendels to study the destruction of the AF state versus hole doping [52], while the $^{89}Y$ NMR permitted to probe the metallic phases beyond the AF state, that I shall describe here-after.

### 4.2 $^{89}Y$ NMR and the pseudogap

Let me recall that even the basic features of the electronic structure of these materials were under question at that time. The respective roles of the Cu holes responsible for the local moments yielding the AF state of the parent compounds and of the doped holes expected to be located on the oxygen orbitals had to be clarified. The macroscopic magnetic susceptibility should sum up the contributions of these two types of holes. NMR experiments could permit to solve that question as the $^{63}$Cu NMR should probe the spin contribution on the copper sites, while the $^{89}$Y nuclear spins should be more likely coupled to the oxygen holes. This encouraged me to consider that the comparison of $^{89}$Y and $^{63}$Cu NMR would be essential. While it has been important so far to underline the context in which the experiments were achieved, I shall now recall some of the results which are not familiar with most researchers in the field.

### 4.2.1 Hyperfine couplings and electronic structure

One had to determine beforehand the actual hyperfine couplings of the nuclei with the different types of holes. Let me recall that the $^{89}$Y NMR shift $^{89}K$ is expected to be given by

$$^{89}K = \delta + K_s = \delta + A\chi_s. \qquad (12)$$

Here δ should be a chemical shift due to the filled molecular shells. I anticipated that the spin contribution $K_s$ would be associated with the spin susceptibility $\chi_s$ of the doped holes, and that its magnitude would be determined by that of $A$, the hyperfine coupling between the $^{89}$Y nucleus and the oxygen hole orbital.

a) *Sign of the hyperfine coupling.*

So the first contribution I could do on YBa$_2$Cu$_3$O$_{6+x}$ was to use the room $T$ access in the magnet to probe at 300K the variation of the $^{89}$Y NMR shift with doping, that is with varying $x$, the chain oxygen content. I did such measurements initially on non-oriented powders. This happened to be a good choice as I did detect narrow $^{89}$Y lines which did not display any anisotropy (that is $^{89}K_c \sim {}^{89}K_{ab}$) and which exhibited a rather large *increase* of $^{89}K$ with decreasing doping down to the AF state [53] (see Fig. 8 where the data including $T$ dependences taken later are reported).

While the macroscopic susceptibility $\chi_m$ and density of carriers were known to decrease with decreasing doping, the corresponding *increase* of the shift proved that the transferred hyperfine coupling $A$ on the $^{89}$Y nuclei is negative. This by itself appeared as an important result, as it excludes an overlap of the oxygen holes with the Y *4s* orbital, which gives a *positive* hyperfine coupling. One can immediately see that this excluded a hole occupancy of the O $2p$-π orbital, which points directly towards the Y and would yield a large overlap with the Y 4s orbital.

This result permitted to resolve [54] an issue which was debated between researchers doing tight binding or LDA calculations and those doing cluster calculations. It ascertained that bands cutting the Fermi level resulted from the hybridization of Cu $3d_{x2-y2}$ and O2$p\sigma$ orbitals, as is totally accepted nowadays.

b) *About the AF state.*

We were able in that experiment [53] to detect the $^{89}$Y NMR signal even in the parent compound YBa$_2$Cu$_3$O$_6$ which displays a Néel AF order below $T_N$ =410 K. The $^{89}$Y NMR shift was found independent of $x$ within experimental accuracy below $x$=0.4. This is easily understood, as the yttrium is located at a symmetric site of the Néel AF lattice, so that the $^{89}$Y nuclei are more sensitive to the metallic band than to the AF contributions.

A full quantitative analysis of the data required a determination of the chemical shift δ. As the hole doping becomes negligible in the AF state we assumed initially that the $^{89}$Y nuclei do not sense the Cu paramagnetism, so that the NMR shift measured in the AF state gave a lower limit δ=22 ppm. The actual value was found larger when the $T$ dependence of $^{89}K$ allowed a better determination..

This work got recognized immediately before publication and the abstract was selected as an oral talk at the Interlaken M2S-HTSC conference in February 1988. The published paper had triggered interest and debates about the sign of the hyperfine coupling [54]. I have then been invited to give a talk in a special invited symposium on NMR in cuprates at the APS March meeting in 1989 (I have many reasons to think today that C. Slichter proposed this symposium and selected the invited speakers).

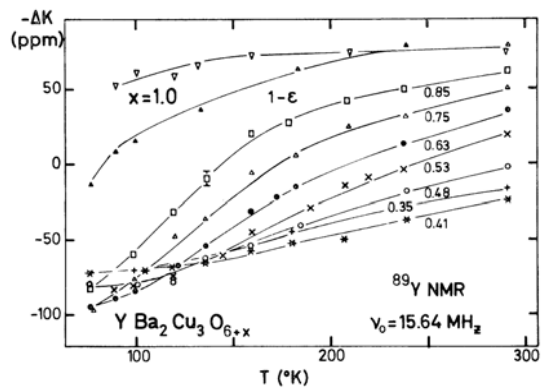

**Fig. 8** Temperature variation of the $^{89}$Y NMR shift –Δ$K$ for YBa$_2$Cu$_3$O$_{6+x}$ powder samples from optimal doping to a non-superconducting sample for $x$=0.41. The progressive increase of the pseudogap magnitude is apparent (from Ref. [55])

### 4.2.2 $T$ dependence of the spin susceptibility and single spin fluid response

As these experiments appeared successful I made a choice of cryostat which I ordered immediately and designed a probe permitting to investigate the $T$ dependences and to reach the SC state. As soon as it became operational, I found a surprisingly large $T$ variation of $^{89}K$ for an underdoped sample with $T_c$=60 K. This was so puzzling that I thought that a technical problem occurred with the brand new probe and cryostat. A thorough control permitted to ensure that the data were real, and we rushed then to take systematic shift and $T_1$ data for various oxygen contents. This was done with the help of T. Ohno, a visiting fellow from Japan. We had to take all possible data before the end of January 1989, as both the magnet cryostat and I had to undergo significant surgeries.

The magnet cryostat had to be exchanged with a new one with less $^4$He and N$_2$ consumptions, while my surgeons had to take off and analyse a throat-tumor. Both interventions were somewhat dangerous, but as both the magnet and I are still here today one should guess that everything got through perfectly. I even have had an unexpected free time in the hospital after the surgery to analyse the data we took (no portable PCs at that time, just a small pocket calculator, graph paper, pencils and a rubber), to begin to prepare my talk for the March meeting and to write a paper.

The large $T$ variations (Fig. 8) detected in underdoped samples [55] were unexpected from the existing published macroscopic susceptibility data, because these samples were actually contaminated by impurity phases displaying a large Curie contribution at low $T$. The great usual advantage of the NMR is that it probes the majority phase and is unaffected by these impurity phases. So these data demonstrated that an unusual $T$ variation occurred specifically for underdoped samples, but to understand its magnitude, one had to determine better the reference chemical shift δ and if it possibly depends on doping. Interestingly, S. S. Parkin from IBM Almaden just visited Orsay at that time and gave a seminar in which he indicated that his $\chi_m$ data taken with a SQUID magnetometer displayed quite similar $T$ dependences, and only weak Curie-Weiss contributions.

The plots of K versus Parkin's $\chi_m$ data displayed in Fig. 9 were linear, with slopes quasi-independent of hole doping, even for a sample with $x=0.45$, which did not display any SC, being at the boundary between the AF and SC regimes. I was immediately aware that this doping independence of the hyperfine coupling constant $A$ revealed an important feature of the electronic structure. Indeed, I expected that the Y nuclei would probe the oxygen holes while $\chi_m$ would be dominated by the Cu electronic spin susceptibility. We rather found that both probe the same $\chi_s(T)$, except for the translation of the curves in Fig. 9, which could be attributed to a $T$-independent contribution either to $^{89}K$ or to $\chi_m$. These results established that the $T$-dependent part of $^{89}K(T)$ is dominated by the Cu hole contribution $\chi_s(T)$, and that $A$ is a transferred hyperfine coupling from Cu through Cu–O2pσ covalence. In other words the Cu holes slightly extend on the O2pσ orbitals by covalent bonding even in the AF parent. The extra $T$-independent contribution to $^{89}K$ or $\chi_m$ could be attributed to variations with doping of δ or of the Cu orbital susceptibility. In any case any specific contribution of doped holes to $\chi_s$ would appear extremely weak. That was on line with the Zhang and Rice [56] suggestion that oxygen holes just form singlets with Cu and only modify the Cu susceptibility so that the Cu and O holes are highly correlated. Though some NMR colleagues suggested that a very weak spin shift due to an independent oxygen band could not be excluded [57], accurate comparisons, to be recalled below, with $^{63}Cu$ and $^{17}O$ data later confirmed the validity of this analysis.

As for the large $T$ dependence, it indicates that the susceptibility of Cu decreases well below the level expected for the AF state of a 2D Heisenberg magnet. Furthermore, the reduction of $\chi_s$ expected from the opening of a superconducting gap at $T_c$ becomes very small in the underdoped samples, as if a partial gap has already been opened progressively with decreasing $T$.

### 4.2.3 Dynamic spin susceptibility and metallic-like component

The other feature that we clearly evidenced in Ref. [55] was that $(T_1T)^{-1}$ and $^{89}K$ have very similar $T$ variations with $T$ and doping. This is illustrated in Fig. 10 (a) on better field-aligned YBCO samples [58] realized later for two compositions, $O_7$ for which $(T_1T)^{-1}$ is $T$ independent as is $^{89}K$, while for $O_{6.6}$ both quantities exhibit large $T$ increases. So the dynamic susceptibility appeared quite correlated with the static susceptibility. I found that this could be a way to determine the value of the chemical shift δ. Indeed, assuming that the Korringa law $T_1T K_s^2$ =cste applies, a rather good fit with the data was obtained [55] for δ = 300ppm, significantly different from that assuming that $K_s=0$ in the AF state. I shall only briefly anticipate that further studies, to be recalled later in §4.2.4, permitted to establish that the constant quantity is rather $T_1T K_s$, once a more reliable value δ = 150 ppm has been derived.

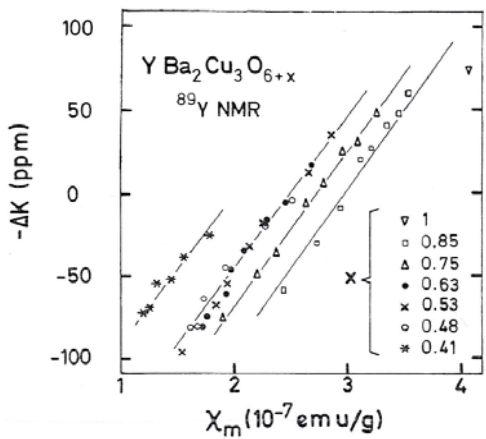

**Fig. 9** The $^{89}Y$ NMR shift $-\Delta K$ for $YBa_2Cu_3O_{6+x}$ from Fig. 8 are plotted versus the macroscopic susceptibility data $\chi_m$ taken by S.S. Parkin for samples with similar oxygen content rather free of parasitic impurity phases. Linear least square fits give parallel slopes for the various samples, that is the same negative value for the hyperfine coupling $A$ (from Ref. [55]).

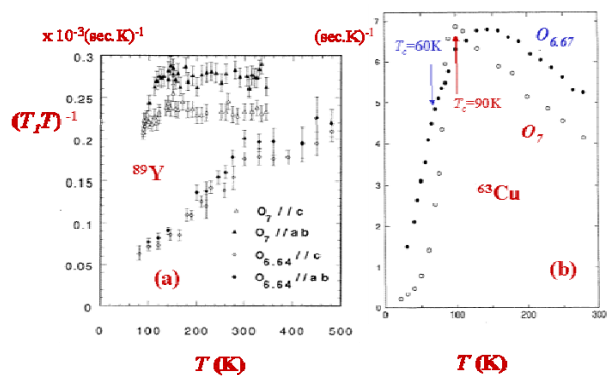

**Fig. 10** Variations with temperature of $(T_1T)^{-1}$ taken for oriented powder samples of YBCO$_7$ ($T_c$=90K) and YBCO$_{6.6}$ ($T_c$=60 K). (a) $^{89}Y$ data taken from ref (58). (b) $^{63}Cu$ data taken from ref (62). While the data for $T_c$=90 K are $T$ independent for $^{89}Y$, as does the Knight shift, they increase at low $T$ for $^{63}Cu$. Similarly for the underdoped samples both $(T_1T)^{-1}$ and $K$ for $^{89}Y$ increase regularly up to $T^* \sim 350$ K, while those for $^{63}Cu$ display a maximum at $T \sim 150$ K, assigned to a spin gap.

### 4.2.4 About the origin of the "pseudogap" denomination and J. Friedel's implication

Of course, I discussed of these data with various colleagues at Orsay when back from the hospital, as I was eager to find an explanation for this large reduction of $^{89}K$, which appeared as a drop in the DOS at the Fermi level, occurring well above $T_c$. I noticed a large spectrum of reactions, from a total lack of interest from D. Jérome to enthusiastic remarks by M. Héritier or H. Schulz. But no obvious interpretation went out of such discussions, although most of us were convinced that this was linked with the existence of correlations between the Cu holes and had to be connected to Mott physics.

I had as well a short discussion with J. Friedel and underlined him that this metallic like behaviour interleaved with correlations was a surprising aspect in the data. He suggested a magnetic analogy with a proposal done by N. Mott for short-range order in alloys. The gap which should open if an ordered state were achieved would not be fully formed in a quasi-ordered state. This would result in a "pseudogap" which would be seen as a drop of the DOS at the Fermi level. This idea was quite sensible, as usual with J. Friedel, but being more influenced by the strong correlation ideas, I did not see how this band picture would be compatible with the Zhang and Rice singlet which appeared to explain the absence of contribution of doped holes to $\chi_s$. In any case I liked the notion of "pseudogap" to qualify the drop of the DOS and decided to use it.

I did not have time to pursue these exchanges with my colleagues as I had to leave for the March meeting in St Louis, where I intended to report at least part of these results and not to restrict my presentation to that of the 1988 paper. The NMR session was given there in a hall with an audience of about 500, so I felt that I could disclose unpublished results. This was not uncommon at that time, as understanding the physics was still the main issue. In any case I knew how much time was needed to repeat our data and that I would submit a manuscript in a short-while.

I think that this amount of information allowed me to publicize the capabilities of NMR, which was the aim searched by C. Slichter. I also participated there to an exciting NMR session of contributed papers on cuprates where we had interesting exchanges on the data on the $^{63}$Cu and $^{17}$O NMR in YBCO$_7$ taken by M. Takigawa and C. Hammel at Los Alamos [60]. They evidenced that $^{63}$Cu and $^{17}$O displayed $T$-independent shifts but that the $(T_1T)^{-1}$ were quite distinct, that on $^{17}$O being $T$ independent as found for $^{89}$Y, while for $^{63}$Cu it increased at low $T$, as shown in Fig. 10(b).

This was then as well a proof that even in the optimally doped samples, for which a pseudogap was not detected, the electronic correlations played a role. The local dynamic AF correlations induced a fluctuating field on the Cu sites, which cancelled on the oxygen and yttrium sites as one could anticipate.

Back in Orsay, I completed the redaction of the manuscript, in which I emphasized that the single spin-fluid picture applied. I put there some emphasis on the metallic-ike behaviour probed by the $^{89}$Y NMR, attributing it to a Fermi-liquid component, although this was maybe a little bit overstated, as I explained at the same time that important correlations were required to explain the absence of independent oxygen band due to the doped holes. For the drop of the DOS, as at St Louis, I used "pseudogap" to qualify it and indicated that this denomination resulted from a discussion with J. Friedel, but refrained to detail any possible explanation for its origin. I had noticed that in his Interlaken paper J. Friedel had already used this term in a different context [47]. He was suggesting there that in 2D systems, SC fluctuations would occur in a large $T$ range above the 3D superconducting transition, and might be seen in the excitations above $T_c$ as a "pseudogap" reminiscent of the SC gap. One obviously needed more work to decide whether the pseudogap was due to magnetic correlations or was a precursor to superconductivity.

As J. Friedel was initially interested in these results, I thought that we would complete the experiments and put together with the Orsay theory group a good explanation for the pseudogap. However at the end of 1989 and in 1990 many debates opened within the cuprate NMR community concerning the pseudogap, and we had to face many questions that I shall discuss briefly in § 4.3. Furthermore, as far as I can recollect, I did not have other opportunities to discuss physics with J. Friedel for a while. From what I can see in our publication record after 1989 the acknowledgements for theoretical discussions were then directed towards M. Héritier or H. Schultz, the latest outstanding student of J. Friedel. These theoreticians were obviously more interested in the strong correlation limits.

This period after 1989 corresponds to the time where J. Friedel had retired from his teaching position and nearly disappeared from the LPS, although he had still his office there for a decade. I expected that he would become more active in research, especially on the HTSC problem, but he did not do that choice. Only J. Friedel himself would be able to give his reasons, but I guess that he initially wished to get somewhat disconnected from the LPS and to let us develop ourselves independently. He took important responsibilities at the Science Academy, that he presided in 1994, and maintained his efforts to orient from there the science policy of the country during the 1990's. He also wrote at that time the biography [2] mentioned in the introduction.

His absence from the LPS coincided with a profound change of his scientific information channels. J. Friedel used to be a member of most PhD committees on condensed matter in France. Having attended many of those, it was obvious for me that this was where he used to be efficiently acquainted with the recent scientific developments. It was remarkable to see him as a president of the jury claiming, after every other member had given remarks and raised questions, "I have nothing to add to all what has been said", and then to ask "minor" questions which always appeared as essential and scientifically pertinent, and which confirmed that he had read the manuscript in great detail.

So being less invited to participate to PhD committees, he lost this information channel. Furthermore, the beginning of the 1990's has been the boom of the use of personal computers and the large advent of information technologies in the scientific community. From what he

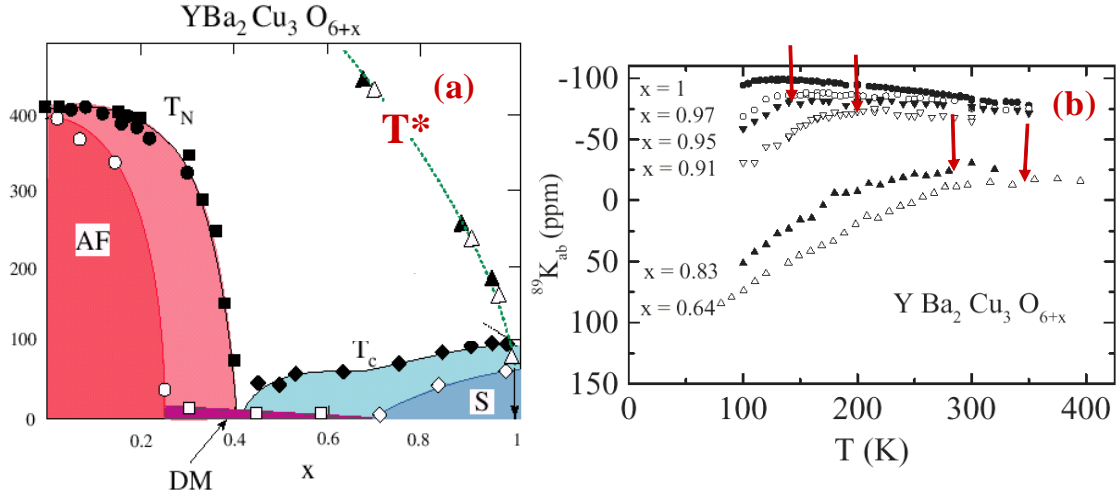

**Fig. 11** (a) Phase diagram versus oxygen content for pure $YBa_2Cu_3O_{6+x}$ (full symbols) and for 3% Zn substitution (empty symbols). The Zn induces a shrinking of the AF and SC phases with the appearance of an intermediate spin glass phase. (b) $^{89}Y$ NMR shift data, more accurate than the initial one of Fig. 8, are presented to illustrate our definition of $T^*$ (arrows). The $T^*$ data in (a), taken on pure and substituted samples prepared with the same heat treatment, are found independent of disorder.

told me recently, J. Friedel did not adopt those and was certainly cut then of the huge flow of information specifically on the HTSC. He also lost as well the benefit of the numerous preprints he used to get from all over the world which disappeared totally from the scientific exchanges.

I recalled earlier that J. Friedel was privileging a band picture with weak correlations to explain HTSC at least at optimum doping. So he was inclined to consider that the correlations which develop in the pseudogap region are not important phenomena for SC. Had he emphasized that these results obtained at LPS opened an unprecedented possibility to study doped Mott insulators, our results would certainly have been highlighted by at least one French scientific institution. On the contrary, in the international HTSC community, the pseudogap was becoming the rising star, and our results got publicized by various physicists, among others C. Slichter, P. Lee and C. Varma, who recently contributed awarding me an APS fellowship in 2007 for work exclusively performed at the LPS[a].

### 4.3 The long lasting pseudogap problem

Meanwhile, the quantitative analysis of data taken by Takigawa and Hammel on $YBa_2Cu_3O_7$ [60] had been performed by Mila and Rice [61], using quantum chemistry estimates of the hyperfine couplings and confirmed that the low T increase of $(T_1T)^{-1}$ of $^{63}Cu$ just evidences that these nuclei probe the magnetic fluctuations at the AF wave vector $Q_{AF} = (\pi,\pi)$.

---

[a] The fellowship has been awarded « For nuclear magnetic resonance studies of strongly correlated electronic materials including the pseudogap phase of the cuprates through Knight shift measurements, local magnetic moments in cuprates, and studies of Kondo effect and spin-glasses ».

Our work on $^{89}Y$ had evidenced that a pseudogap temperature can be pointed at the onset $T^*$ of the decrease of the NMR shift magnitude. For decreasing doping, this $T^*$ and then the pseudogap width, increase sharply with decreasing doping [55] (Fig. 8).

Stimulated by our results, M. Takigawa and C. Hammel repeated their work on $^{17}O$ and $^{63}Cu$ on an underdoped sample with $T_c$=60 K and established that their NMR shifts displayed the same $T$ dependence as that of $^{89}Y$ [62]. This confirmed the single spin-fluid picture and the scaling of the data for the three nuclei permitted to determine δ = 135-150 ppm for the $^{89}Y$ NMR. However, as was seen as well by others [63-64], they found that the $(T_1T)^{-1}$ of $^{63}Cu$ goes through a maximum in this underdoped sample at a temperature much lower than $T^*$ (see Fig.10 b). This could be interpreted as indication for a precursor pairing [63], or the opening of a "spin gap" at $Q_{AF}$, the AF wave vector, as proposed by H. Yasuoka at a workshop in Japan [64]. Sometimes labelled as "spin pseudogap", its occurrence at $Q_{AF}$ has been confirmed by inelastic neutron scattering experiments [65], and the temperature of the maximum was found to increase much less rapidly than $T^*$ for decreasing doping [66].

Most other experiments measuring uniform macroscopic responses, such as specific heat [67], and planar resistivity $\rho_{ab}$ [68], do detect an onset at similar temperatures [69] as that of $T^*$. So this $T^*$ obtained as the onset of the decrease in temperature of the Knight shift is undoubtedly the highest temperature below which a detectable phenomenon occurs. This $T^*$ line is reported in the YBCO phase diagram of Fig 11.

Both the pseudogap $T^*$ and the spin gap are detected only in underdoped samples which suggests that they are connected. These experimental results might have been precursor indications of the **k** space differentiation which has been found by spectroscopic ARPES or STM experiments. However, so far no serious quantitative comparison between these experiments has been attempted.

Finally all the extensive studies that have been performed, reveal the presence of a metallic-ike component, as that we had detected on the $^{89}$Y NMR, which we had labelled as Fermi-liquid like. Phenomenological attempts to describe the spin susceptibility $\chi''(\mathbf{q},\omega)$, have been done [70] by Millis Monien and Pines (MMP). In all cases both this metallic component and the pseudogap had to be introduced by hand, an approach which is not really satisfactory on the theory side.

When coming back to these old experimental results and their analyses, one has to keep in mind that twenty years of intense activity have followed and that a huge amount of data has been taken with improved experimental accuracy and with refined novel techniques. Those are so novel that they have initiated vivid debates on the pseudogap, but did not permit so far to resolve the issues they raised. The pseudogap remains still today the central point debated at any conference on the cuprates.

### 4.3.1 Impurities and phase diagram

To probe the relation between pseudogap and superconductivity, I have immediately launched in 1990 the most naïve experiment one could imagine. We knew that Zn impurity substitutions on the planar Cu sites depress $T_c$ very efficiently. So it was natural to look at their incidence on the pseudogap [71]. The result of this experiment was unambiguous, as we found that the pseudogap T*~350 K observed on the main NMR line was unaffected even if $T_c$ is decreased to zero. This allowed us to suggest that the superconductivity and the pseudogap are independent phenomena.

In the same experiment we got a new surprise, as we found that the non-magnetic Zn impurities induce a local moment like paramagnetism in their surroundings, which is the signature of the response of the correlated electronic state of the host material. The fact that the electronic behaviour is mostly modified in the vicinity of the Zn, is also evidence that the pseudogap is a short-range phenomenon and does not require a long-range ordering to be detected.

We could refine later these studies when we could detect the NMR of the near neighbour of Zn or Li substituted on the planar Cu sites [72-73]. I shall not get into the description of the large amount of work we did over more than 10 years to exploit the underlying information on the spatial extension of the perturbation, which is related to the magnetic correlation length of the pure compounds. The basic idea which has been governing our effort is that impurities always reveal the properties of the pure material. An interested reader could find those in our review paper [74] on these impurity effects in correlated electron systems.

### 4.3.2 Generic and material specific properties

But so far these observations of the pseudogap were limited to YBCO, as a clear decrease of the DOS could not be detected by NMR in the 124 compounds. This had been tentatively attributed to a difference between single-layer and bilayer systems. We therefore searched for a distinct single-layer compound. Our colleagues J.F. Marucco and D. Colson at SPEC/CEA Saclay were synthesising the Hg1201 compound and controlling its oxygen content. So we could do on such powder samples an $^{17}$O exchange within the PhD of J. Bobroff [75]. In the underdoped phases of this single-layer material the $^{17}$O NMR allowed us to evidence a pseudogap displaying similar $T^*$ values and hole doping dependence as that seen in the bilayer YBCO. This result established that the pseudogap is a generic property of cuprates.

Later on, in the three-layer Bi 2223 system with nearly optimal $T_c$ it has been found that the central $CuO_2$ layer and the external $CuO_2$ layers had different $T$ dependences of the NMR shift [76], suggesting that the external layers are overdoped while the inner one is underdoped. In that sense the occurrence of the pseudogap does then permit to determine the doping of the different layers. This has been applied extensively to various multilayer Hg compounds recently [77].

We suggested altogether [75] that LSCO was not a material displaying generic properties and that its low $T_c$ value could be associated with material specific problems (e.g. disorder due to local stripe order). We demonstrated later that NMR experiments do evidence a larger disorder in the lower $T_c$ compounds than YBCO, especially near the chain ordering compositions YBCO$_7$ and YBCO$_{6.6}$, so that these compounds, as well as the Hg1201, appear to be the cleanest cuprates [78].

We could confirm this as well by performing transport measurements on single crystals in which defects are introduced by electron irradiation at low temperature, which creates point defects such as Cu or O vacancies [79]. There we found that the metal insulator transition (MIT) is fully driven by disorder. We could even correlate the transport behaviour observed on pure lower $T_c$ compounds with that induced by introducing controlled disorder in YBCO [80]. This permitted to conclude that the lower the optimum $T_c$, the higher the concentration of holes for which the MIT appears and the wider the range of hole dopings in which a spin glass regime is detected.

It took about ten years to convince the community that the observations done in the lower $T_c$ compounds are somewhat specific and not generic of the properties of clean cuprates. The doping of the cuprates being obtained by chemical substitution or by insertion of extra oxygen in the intermediate layers between the $CuO_2$ planes, disorder effects are hardly avoidable and do appear to influence most of the physical properties. The low doping side of the phase diagram is the most affected by disorder, which governs the MIT.

Concerning the materials aspects, the synthesis of single crystals is absolutely essential to perform transport, STM, ARPES measurements. Many of our colleagues are highly influenced by sentences such as: "We have performed these measurements on high-quality single crystals". One has to be aware that this HQSC qualifier is meaningless as long as clear quality criteria have not been established! But one should also be aware that a naïve but robust belief in the community is that single crystals are always better than ceramic samples. This is by far not true as single crystals are usually synthesized at high temperature, using flux techniques, etc… which might produce extra defects with respect to solid state synthesis. The phase diagram which was established versus Sr content in doped $La_2CuO_4$ is still quite often taken as a reference phase diagram of the cuprates. We have proposed [80] ([81]) that a disorder axis should be used, as shown hereafter in Fig. 12, and that SC is stable for much lower doping in the cleanest cuprates.

To conclude this section, I would like to stress again here the great advantage of the NMR technique. It remains without doubt one of the best probes allowing one to disclose the local spatial responses in macroscopic samples of the complicated materials which are at the heart of the correlated electron physics. This advantage has quite often been used in Orsay in the various experimental studies performed not only on HTSC but more recently on cobaltates [82], fullerites [83] or pnictides [84].

### 4.3.3 Is the pseudogap due to preformed pairs or to a competing order?

In the early NMR experiments depicted above the pseudogap has been shown to be generic to underdoped cuprates and to be robust to disorder. However its actual origin was beyond reach even though most NMR specialists attributed it implicitly to Mott physics rather than to a preformed pairing scenario. However strong supports for the latter have been advocated from the detection of a large Nernst effect and of diamagnetism [85-86] above $T_c$. Indeed, such effects could be associated with superconducting fluctuations (SCF) and/or vortices persisting in the normal state. Therefore many experiments have been devoted to attempt to locate a new crossover line in the phase diagrams of Fig. 12 at temperatures $T'_c$ above which SCF become undetectable.

Using high magnetic fields to suppress the SCF to the conductivity [87], we could determine precisely altogether the onset temperatures $T'_c$ of the SCF and of the pseudogap $T^*$ within the same set of transport experiments [88]. In a further extensive quantitative study [81] of the SCF we have demonstrated that $T'_c$ can be reliably considered as the onset of pair formation and that $T^*$ occurs below $T'_c$ at optimal doping. As indicated in Fig. 12, the pseudogap line $T^*$ crosses the onset of SCF, which evidences unambiguously that the pseudogap cannot be a precursor state for superconducting pairing. The influence of disorder on SCF has been thoroughly studied as well in [81,88], and we have shown that the local pair formation underlined by the $T'_c$ line is only moderately affected, while the bulk $T_c$, that is the SC pair coherence can be severely reduced by disorder. Our results allow us to draw important conclusions on the incidence of disorder on the phase diagram of cuprates, which are sketched in Fig. 12.

While the preformed pair scenario has still some indefectible supporters, there are growing evidences which fit with our results and the views presented above. Recent new experiments reveal the existence of competing orders of magnetic origin, with possibly a cascade of phases developing towards lower $T$. Extremely controversial points of views are expressed about these electronic properties in the pseudogap regime. Even a simple realistic enumeration of those would represent a huge work that is not really the aim of this report. But this absence of consensus shows that the cuprates and the pseudogap have still beautiful days of exciting life ahead.

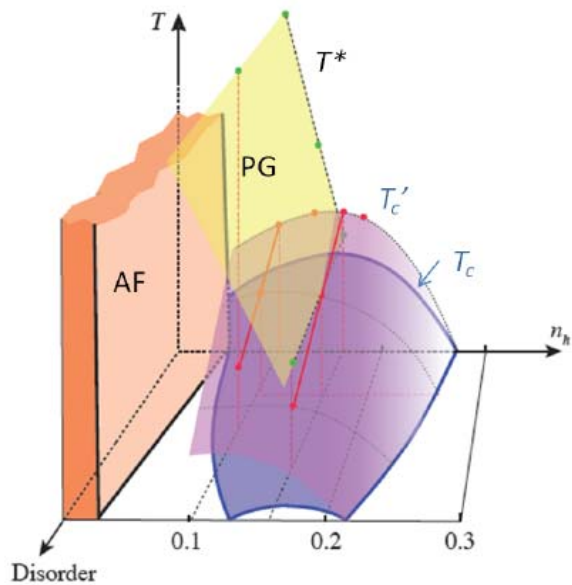

**Fig. 12** Phase diagram showing the evolutions of the onsets of the pseudogap $T^*$ and of the SC fluctuations $T'_c$ with doping and disorder. The pseudogap and the SCF surfaces intersect each other near optimum doping in the clean limit These surfaces have been limited to experimental ranges where they have been determined experimentally. We did not represent any putative evolution of $T^*$ below $T_c$, nor of $T'_c$ towards the regimes where static spin-glass-like magnetic disorder appears. In the overdoped regime, data taken on Tl2201 [89] indicates that disorder suppresses SC without any anomalous extension of the SCF (figure from ref. [81]).

# 5 About the underlying evolution of our scientific community

In this essay I have tried to illustrate two distinct scientific periods separated by the discovery of HTSC (and the advent of information technologies). I got my education from "elephants" born before the 2nd World War. At that time scientists appeared interested in solving and understanding problems in fair scientific confrontations which were dominated by Bell Laboratories. I then illustrated how things changed in the 1990s after the discovery of HTSC. I shall try here to complement this comparison by aspects which are in my opinion quite important for the future of this scientific research field that is materials, modelling and theoretical understanding, and incidence of publication policies and finally scientific education.

## 5.1 Materials

I have recalled here that materials have always been at the heart of the important problems encountered in solid-state physics. This becomes a really dominant aspect in the recent evolution of correlated electron physics which addresses more and more complicated materials. This aspect was well anticipated by J. Friedel when he created the LPS. This situation was as well encountered in labs connected with industry, such as Bell Laboratories. It remains however very difficult to associate distinct scientific communities within the university systems in which materials approaches are not necessarily well highlighted. But the importance of collaboration between material science and physical measurements has been illustrated in HTSC physics as most new results have been obtained thanks to careful control of materials. This has been our case in Orsay, also in Zurich, Japan, Vancouver and now in China for pnictides. For the future of condensed matter physics, one needs to reward collaboration between disciplines and interpenetration of the measurement techniques. This evolution is essential but in my opinion generates some difficulties in the relation between experiment and theory as discussed below.

## 5.2 From experiment to theoretical understanding

The role of experiments is to help to isolate the pertinent parameters which drive an observed phenomenon and to highlight a minimal theoretical model. These steps, in which J. Friedel excels, require phenomenological approaches with possibly good guesses and the ability to integrate all the existing experimental information. For the Kondo case this was possible and was done by experimentalists. The experiments did not evidence large differences in behaviour for local moments of different magnitude, so that a single-orbital exchange coupling model was accepted in a few years as generating the basic Kondo phenomenon. Accurate theoretical solutions were required at a later stage of course and were available in less than 10 years in that case.

The situation for HTSC is much less obvious, and after 25 years the phenomenology has not been sorted out, as the incidence of materials is important and makes difficult the link between models and real materials. As a result, the model Hamiltonian is not so clear for cuprates. Would it be sufficient to solve the doped Mott-Hubbard insulator, the t-J model, or is a three-band model required? Even the choice of the starting model appears difficult, as one has to distinguish first the relevant generic properties.

So theorists are hardly encouraged to attempt difficult theoretical calculations within a given model. Only the Mott transition for the undoped Hubbard model has been solved correctly so far using the new DMFT approaches. Actual theoretical developments being difficult, theorists have been more akin to perform themselves the phenomenological analyses, while they are not qualified to study material problems that they tend usually to consider as marginal. Moreover, with the advent of new techniques, which are not always well understood either, less and less researchers can map together information coming from different experiments.

This has led in the case of HTSC to a somewhat stable organisation into distinct "quasi-religious sects" studying models of various types of orders which might explain the pseudogap with the underlying speculation that those models might also describe all the properties of the cuprates. Though these problems are often interesting per se these speculations might be wrong. So doing, theorists of different chapels remain key players who govern the field by highlighting a limited set of experiments, if not a single one, which they might explain by some ansatz.

While the experimental work becomes much harder and more refined, the advent of new techniques like optical, Raman, ARPES and STM, which are energy and **k**-dependent spectroscopies, is extremely important but with technical limitations in resolution. One has also to decide whether the surface properties are representative of those of the bulk, so that the impact of such techniques might be sometimes overemphasized. ARPES has an important role as it helps at least to find out differences with LDA. This technique therefore is quite important to help linking band structure computations with the experimental situation and to determine the importance of electronic correlations.

These remarks apply of course not only to HTSC but also for most novel materials like layered pnictides, chalcogenides or cobaltates, on which the community has been quite active recently. Overall, this also shows how important are the information channels that I shall consider below.

## 5.3 Incidence of the publication policies on the careers of researchers

I have underlined that the discovery of HTSC has been concomitant with the expansion of information technologies. Nowadays the fastest vehicle of scientific information in our field are the on line *condmat archives*, which display at least 50 papers per day. This replaces quite efficiently the mailed preprint system which was limited towards selected individuals. For topics which involve only a small scientific community, this is extremely efficient. But for broad topics such as HTSC, the accumulated amount of information is huge, and might kill information. Few scientists have enough time and are competent enough to dress a hierarchy of the existing information which has been accumulated for more than 20 years.

This has completely changed the role of journals and initiated a large evolution of the publication policies. In the past, scientific information was dedicated to specialized scientists, and Physical Review Letters (PRL) was ensuring in principle fast publication and selectivity by referees. Though this system was often criticized, nobody could start a better one, and PRL remained a reference. Within the scientific community, journals like Nature or Science were considered as magazines done to popularize limited aspects of science and not as journals dedicated to ease the exchanges between scientists.

Meanwhile a constant reproach had been done to scientists, especially in our field of research, as we did not do significant efforts to popularize our science towards larger scientific communities. This was not the case for particle physicists who were involved in large consortia driving huge experiments and were efficiently represented by professional scientists specialized in outreach. In our field of research individual players with single authored articles were not uncommon, but few were ready to lose their time defending the field and improving the life of the community. Nature and Science journals could have played at least partly this role, as their editors were free to select and publish work which could impact any readers out of the given research field. The big surprise for me is that knowing the partiality of the editorial system, our scientific community has progressively accepted to highlight excessively the incidence of publications in these journals. Even the Schön-Battlogg case did not put an end to this up-rise, and although many scientists privately criticize this evolution, they all try to publish there in priority.

I would say that PRL has even helped that evolution by trying to follow similar rules. PRL lost the fast publication argument to select the accepted papers, after the publications in the archives took the leadership. In my opinion, the APS made the great mistake to choose to reduce the volume of PRL and to try to mimic the methods of Nature or Science. Most journals nowadays want to favour papers accessible to a large audience of physicists in order to get an increase of their impact factors. Mixing popularization of science and science itself might not be the best choice.

This evolution of the community has validated what I call the "physics of scoops". Following other trends of the society, research teams are now transformed into soccer teams trying to mark goals against other teams. For a PhD student or a post-doc who has to find a job, understanding the physics becomes much less essential than cumulating goals possibly obtained by jumping from one research group to another. The more goals you mark, the larger chances to get a job. So (i) collective work is over emphasized, with papers frequently signed by 10 to 20 authors (ii) students select their post-doc positions by counting the number of publications in Nature and Science of the putative hosting groups (iii) research groups cultivate their relations with specific editors of Nature or Science, and (iv) one can mark goals with overstatements, if not wrong or dishonest assertions, without real penalty. In HTSC everything and its contrary has been "demonstrated" so far, sometimes by the same researchers.

All this is naturally emphasized by the extensive use of bibliometrics. When the H factor was established, it was aimed at measuring the past publications of scientists. Now, knowing this index beforehand biases the citation and publication policies at least of researchers looking for jobs, and this will have large sociological implications in the future. The editors of Nature, usually PhD's who did not make it up to a tenure track position, are indeed playing a major role in deciding of the future of the universities, by selecting the researchers that will get tenure tracks. I might exaggerate a little bit, but if the scientific community does not get a step back, this will be true at least at the average level. The scientific community has to react to decide if it wants to protect its true values!

No doubt that this system has no relation with the way science proceeded at the time I started my PhD and that J. Friedel would not feel at ease in such a system. That might explain why he left aside the information technologies and continues to exchange nowadays by handwritten letters.

### 5.4 Science education in condensed matter physics

All this has of course large incidences on teaching and science education. My own experience has been totally obtained at Ecole Polytechnique where I have been teaching since 1980. This allowed me to be in contact with physicists of different specialties. I have been elected as a professor in 1996. This is for me the opportunity to thank here J. Friedel who recommended me to the selection committee.

Since my nomination, I have been teaching there at the undergraduate level in courses which were not compulsory. But at that time most students were fighting to get accepted for graduate studies driven towards mathematics for finance. So the number of physics students declined regularly, and those who remained were more often interested in theoretical physics. The situation which prevailed during my own studies, which lead me to become a scientist and to get the opportunity to exert that passion continuously would hardly occur nowadays due to the difficulties to get physics jobs.

I think that scientists can hardly fight against money makers, and our only way to attract youngsters is seduction. Those who are attracted by theoretical physics might still shift later towards condensed matter. Concerning the experimental approaches, we need to expose the students to the richness of the field and to differentiate ourselves from mathematical physics. I have tried to do that at the undergraduate level and to show the dominant role of experiment. We have to show to the students how one can learn from experimental observation and that experiments are not here only to verify theoretical prediction (Fg. 13). In our field one can run experiments, react to the results and design new ones in a short-while.

I wrote lecture notes in which I have done an effort to put together formal treatments and the philosophy of observation and created with my colleagues a set of problems and questions which are dedicated to enlighten the scientific approach. I am not sure that this is sufficient to overcome the career financial handicaps, but I think that this might leave some positive traces even for students who will not pursue a research career. This approach might become more efficient now that many foreign students better acquainted with the university system are entering that school and that lectures are given in English. This has justified translating my lecture notes in English [48] recently so that they are accessible to the international community. If not efficient in the French education system, such approaches might be beneficial for students from emerging countries where research remains a vector of social promotion.

## 6 Conclusion

To conclude, I have tried in this essay to give a short scientific description of two important problems on magnetism and SC in which I have been deeply involved at two periods of my scientific activity. While writing that, I have myself realized the importance of the spirit introduced by J. Friedel at the LPS in the positive output of my own work. This text has been written in a short while and only reports personal recollections of events which occurred while I was chained to my experiments and struggling to get scientific recognition. There is therefore no pretention for any reliable historical content in this report which should at most be considered as a testimony, at a given time, on the way I feel that research and education in our field of research has been evolving.

I hope that this will at least trigger some reactions from the readers, and primarily from J. Friedel. Most of the researchers who are today below 50 years old have been educated within the very enthusiastic eras that our field have experienced recently, that of HTSC, pnictides, nanotubes or graphene,… They have been extensively exposed as well to information technologies and recent publication policies. I hope that some of them will keep in mind the importance of intellectually driven research activities which are at the heart of the historical values of our field of research and of the education system which it conveys.

## Acknowledgements

I would like to acknowledge here J. Friedel and all my former professors, who helped to elaborate this nice institution in which I have spent the largest part of my life. The technicians, engineers or administrative staff, have always understood that the researchers do not count their efforts to satisfy the passion which drives their activity. I would like to thank them all as well as their role is essential. Of course a special thank is directed towards the students, post-docs and colleagues who participated in the activity of my research group. P. Bernier and F. Hippert have been important in the work on Kondo effect. More recently P. Mendels, J. Bobroff and V. Brouet who are presently permanent research staff of the laboratory, have shared with me some enthusiastic periods. I hope that though some of the latter ones did not have the opportunity to be taught directly by J. Friedel, they will better understand the important role he has played in the spirit which remains still lively in the LPS, and that they will be able to continue to favour some of the approaches I have been underlining above.

I have highly appreciated the collaborations I have had at Polytechnique on teaching matters with H. Schulz (who unfortunately left us too early), G. Montambaux and H. Pascard. Of course F. Rullier Albenque has been invaluable for her help in teaching and also for the fruitful collaboration we have had for the last ten years, in which she taught me a lot about transport properties. I would like to thank her for the careful reading of this manuscript, and for her patient encouragements and advises.

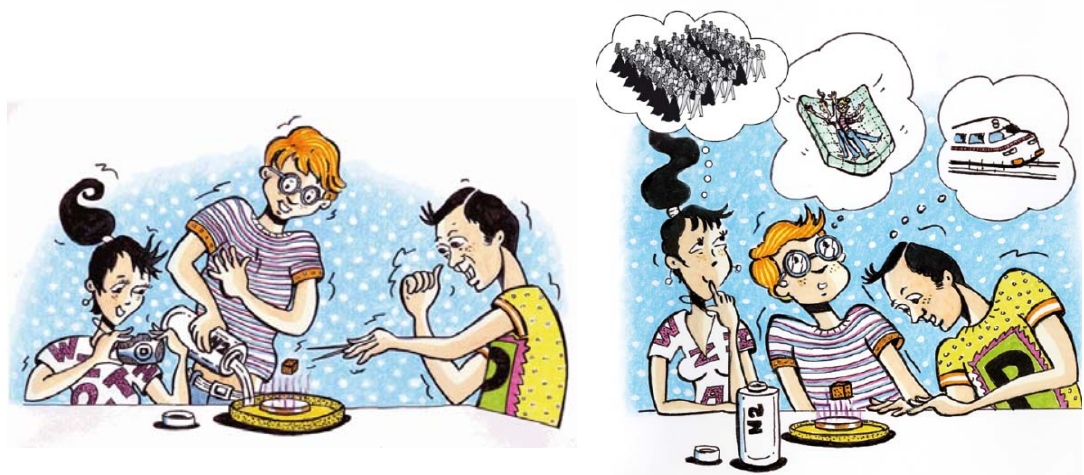

**Fig. 13** Cartoons, taken from Ref. [48], sketching that science involves observation as a first step followed by modelling and/or applied objectives. Only the former intellectual activity is actually specific to the university research system